\documentclass[prd,showpacs,showkeys,nofootinbib,floatfix,preprint,12pt,tightenlines]{revtex4-2}

\usepackage{amsmath,amssymb,revsymb,graphicx,dcolumn}
\usepackage{marginnote}
\usepackage{mathrsfs}
\usepackage{color}
\usepackage{caption} 
\usepackage{subcaption} 

\newcommand{\beq}{\begin{equation}}
\newcommand{\eeq}{\end{equation}}
\newcommand{\beqa}{\begin{eqnarray}}
\newcommand{\eeqa}{\end{eqnarray}}
\newcommand{\bsubeqs}{\begin{subequations}}
\newcommand{\esubeqs}{\end{subequations}}

\newcommand{\dd}{\mathrm{d}} 
\newcommand{\diag}{\ensuremath{\mathrm{diag}}}  

\begin{document}
\title{Shadows and rings of a de Sitter-Schwarzschild black hole}%
\vspace*{1mm}

\author{Zi-Liang Wang}
\email{ziliang.wang@just.edu.cn}
\affiliation{Department of Physics, School of Science, \\
Jiangsu University of Science and Technology, Zhenjiang, 212003, China\\}
\begin{abstract}
\vspace*{1mm}\noindent

We study the optical appearance of a de Sitter-Schwarzschild black hole and its distinguishability from a Schwarzschild black hole. By exploring various accretion models and emission profiles, we investigate the impact of different parameters on the observed shadows and intensity profiles. Our analysis reveals that the outer edge of the shadow, corresponding to the apparent radius of the photon sphere, remains consistent regardless of the spherical accretion details or the size of the black hole. However, subtle differences in the overall brightness and intensity distribution can arise between these two black holes, especially for emission models with sharp peaks near the event horizon. We find that the de Sitter-Schwarzschild black hole tends to exhibit a slightly darker appearance in certain scenarios, while in others, it can appear slightly brighter than the Schwarzschild black hole. These distinctions become more prominent as the radial emission decreases more rapidly. Nevertheless, the size of the shadow alone is not sufficient to differentiate the potential differences in the optical appearance between the de Sitter-Schwarzschild black hole and the Schwarzschild black hole. Instead, distinctions may be observed in the overall brightness of the image. 

\vspace*{-0mm}
\end{abstract}


\maketitle

\newpage

\section{Introduction}
\label{sec:Intro}
Black holes, as predicted by general relativity, are fascinating objects in the universe that arise from the collapse of massive stars or through other astrophysical processes. They are characterized by an extremely strong gravitational field, which is so intense that even light cannot escape from their gravitational pull.  The simplest black hole can be described by the Schwarzschild solution~\cite{schwarzschild1916gravitationsfeld} with a gravitational mass $M$. The Schwarzschild solution has been well tested by many experimental tests, such as the deflection of light and the gravitational redshift~\cite{misner1973gravitation,wald2010general}. However, this solution has a spacetime singularity at $r = 0$~\cite{hawking1973large}, with infinite energy density. The first model of a nonsingular black hole was proposed by Bardeen~\cite{bardeen1968non}, who introduced the concept of a charged matter core within the black hole instead of a singularity. Subsequently, Dymnikova~\cite{Dymnikova:1992ux} proposed a model of nonsingular black hole,  known as the de Sitter-Schwarzschild black hole~\cite{dymnikova1996sitter},  where the Schwarzschild singularity is replaced by a regular de Sitter core. Since then, various models of nonsingular black holes have been proposed and extensively discussed in the scientific literature~\cite{Borde:1996df,Ayon-Beato:1998hmi,Ayon-Beato:1999kuh,Bronnikov:2005gm,Hayward:2005gi,klinkhamer2014new,Frolov:2016pav,Carballo-Rubio:2018pmi,Wang:2022ews}. 

A few years ago, a groundbreaking achievement has been made by the Event Horizon Telescope (EHT) through the observation of an ultra-high-angular-resolution image of the supermassive black hole in M87~\cite{EventHorizonTelescope:2019dse,EventHorizonTelescope:2019uob,EventHorizonTelescope:2019jan,EventHorizonTelescope:2019ths,EventHorizonTelescope:2019pgp,EventHorizonTelescope:2019ggy}. This image revealed a central dark region known as the black hole shadow, encircled by a bright ring. The shadow of the black hole  is a direct consequence of the strong gravitational field, which bends and captures light rays that approach the event horizon. 

The theoretical investigation of the image of a geometrically thin accretion disk around a Schwarzschild black hole was first conducted by Luminet in 1979~\cite{Luminet:1979nyg}. The image of a black hole with a spherical accretion was subsequently studied in Ref.~\cite{Falcke:1999pj}, highlighting the robustness of the shadow's features. Currently,  considerable attention is directed not solely towards the shadows and images of black holes~\cite{Ovgun:2018tua,Gralla:2019xty,Narayan:2019imo,Khodadi:2020jij,Zeng:2020dco,Peng:2020wun,Okyay:2021nnh,Gan:2021xdl,Perlick:2021aok,Wang:2022yvi,Guo:2022nto,Vagnozzi:2022moj,Pantig:2022ely,Pantig:2022gih,Uniyal:2022vdu,Wang:2023vcv,Gao:2023ltr,Nozari:2023flq,Uniyal:2023ahv,Heydari-Fard:2022jdu}, but also towards observing the appearances of other compact objects~\cite{Cardoso:2019rvt,Rosa:2023qcv,Rosa:2023hfm}, such as wormholes~\cite{Bambi:2013nla,Ohgami:2015nra,Wang:2020emr,Huang:2023yqd,Wang:2023rfq} and naked singularities~\cite{Shaikh:2018lcc,Shaikh:2019hbm,Dey:2020bgo,Gyulchev:2020cvo}. These investigations aim to explore and understand the distinctive features of their respective shadows and further our understanding of gravity and the nature of compact objects in the universe.

In this paper, we will investigate the optical appearance of the de Sitter-Schwarzschild black hole (de-SBH)~\cite{Dymnikova:1992ux} when illuminated by different types of accretions. Our main objective is to explore whether it is possible to distinguish between the de-SBH and the Schwarzschild black hole (SBH) based on their respective shadows and rings. The paper is organized as follows. In Section~\ref{sec:Null geodesic}, we  study the null geodesics of the de-SBH. Additionally, the observer's local proper reference frame is also discussed in this section. The optical appearance of the de-SBH, illuminated by accretion from a thin disk,  is examined and compared to the SBH in Section~\ref{sec:Shadows and rings-disk}. In Section~\ref{Optically and geometrically thin spherical emission}, we explore the optical appearance of the de-SBH under both static and radially free infalling spherical accretions. Section~\ref{sec:conclusion} is dedicated to the conclusion and discussion. In Appendix~\ref{sec:Reveiw of de Sitter-Schwarzschild spacetime}, we provide a brief review of the de Sitter-Schwarzschild spacetime. Furthermore, in Appendix~\ref{sec:Geodesics of de-SBH}, we present a systematic analysis on the geodesics of de-SBHs. 

Throughout this paper, we work in reduced-Planckian units with $G = c = \hbar= 1$, where $G$ is Newton's
gravitational constant, $c$ the speed of light in vacuum and $\hbar$ the reduced Planck constant.

\section{Null geodesics of de Sitter-Schwarzschild black holes}
\label{sec:Null geodesic}

In our notation, the metric for the de Sitter-Schwarzschild spacetime is given by~\cite{Dymnikova:1992ux} 
\begin{align}\label{eq:metric_dssBH}
    \dd s^2=-\left[1-\frac{r_g}{r} \left(1-{\rm{e}} ^{-r^3/r_0^2 r_g}\right)\right]\dd t^2+\left[1-\frac{r_g}{r} \left(1-{\rm{e}} ^{-r^3/r_0^2 r_g}\right)\right]^{-1} \dd r^2+r^2\dd \Omega ^2\,,
\end{align}
where $\dd \Omega ^2$ is the metric of a unit sphere. The parameters in metric~\eqref{eq:metric_dssBH} are defined as
\begin{subequations}
\begin{align}
r_0^2&=\frac{3}{\Lambda}\,,\\
  r_g&=2M\,,
\end{align}
\end{subequations}
where $\Lambda$ is the cosmological constant and $M$ is the total mass measured by an observer at infinity. 

The de Sitter-Schwarzschild spacetime \eqref{eq:metric_dssBH} is globally regular,  replacing the singularity of the Schwarzschild spacetime with a regular de Sitter core. For instance, the Kretschmann scalar $ K \equiv R_{\alpha \beta \mu \nu} R^{\alpha \beta \mu \nu}$ reaches a finite value of  $ 24 /r_0^2$ at $r \to 0$~\cite{Dymnikova:1992ux}. At large value of $r$, the solution \eqref{eq:metric_dssBH} coincides with the Schwarzschild metric, with discrepancies only evident at small $r$. This paper aims to investigate whether these discrepancies are reflected in the optical appearance.  

For values of $r_g$ greater than the critical value $r_{g.\rm{cr}}\approx 1.75759 r_0$, the metric \eqref{eq:metric_dssBH} describes a black hole with two event horizons, $r_{h} ^+$ and $r_{h} ^-$.  These two horizons degenerate at the critical value. On the other hand, for $r_g$ values less than $r_{g.\rm{cr}}$, the metric describes a gravitational object without an event horizon. In this paper, we will focus our attention on the case of black holes i.e, $r_g \geq r_{g.\rm{cr}} $. The location of the event horizon for the de-SBH with different ratio $r_g/r_0$ is summarized in Table \ref{table1}. In Appendix~\ref{sec:Reveiw of de Sitter-Schwarzschild spacetime}, we present a brief review  of de Sitter-Schwarzschild spacetime. For more discussion on the Sitter-Schwarzschild spacetime, see Ref.~\cite{Dymnikova:1992ux}.

Due to the spherically symmetric of the metric, we could confine the photon motion on the equatorial plane ($\theta=\pi/2$) without loss of generality. Then, by defining two conserved quantities along the geodesic (with affine parameter $\lambda$)
\begin{subequations}
    \begin{align}
        E&=\left[1-\frac{r_g}{r} \left(1-{\rm{e}} ^{-r^3/r_0^2 r_g}\right)\right] \frac{\dd t}{\dd \lambda}\,,\\
        J&=r^2 \frac{\dd \phi}{\dd \lambda}\,,\label{eq:def_J}
    \end{align} 
\end{subequations}
the radial equation of motion for photon can be written as
\begin{align} \label{eq:r-geodesic}
    \left( \frac{\dd r}{\dd  \lambda}\right)^2 +2V_{\rm{eff-null}}=E^2\,,
\end{align}
with the effective potential given by 
\begin{align}\label{eq:effeV}
    V_{\rm eff-null} =& \frac{J^2}{2r^2}-\frac{J^2r_g}{2r^3}+\frac{J^2r_g}{2r^3}{\rm{e}} ^{-r^3/r_0^2 r_g}\,.
\end{align}
For details on the derivation of Eqs.~\eqref{eq:r-geodesic} and \eqref{eq:effeV}, see Appendix \ref{sec:Geodesics of de-SBH}.  

In general, photon spheres exist at the maximum of the effective potential
\begin{align}\label{eq:V_max}
    \frac{\partial V_{\rm{eff-null}}}{\partial r}\Big| _{r=r_{sp}}=0\,.
  \end{align}
As shown in Table \ref{table1}, the location of the photon sphere for de-SBHs is nearly identical to that for SBH. 

Combining Eqs.~\eqref{eq:def_J} and \eqref{eq:r-geodesic}, we obtain 
\begin{align}\label{eq:photon_orbit}
   \left( \frac{\dd r}{\dd  \phi}\right)^2-\frac{r^4}{b^2}+r^2-r_g r\left(1-{\rm{e}} ^{-r^3/r_0^2 r_g}\right)=0\,,
\end{align}
where $b\equiv J/E$ is identified as the impact parameter for null geodesics that reach infinity (this interpretation will be explained later). The solutions for Eq.~\eqref{eq:photon_orbit} give the trajectories of photon in the vicinity of a de-SBH.

\begin{table}
    \begin{tabular}{|c|c|c|c|c|}
    \hline 
    & & & & \\[-6pt]
    $r_g/r_0$&$r^{+}_{h}/M$ &$r^{-}_{h}/M$&$r_{{sp}}/M$ &$b_c/M$ \\
    \hline
    $1.76$&1.7386 &1.6631&2.9990 & 5.1960  \\
    \hline 
    $1.82$&1.8618 &1.4860&2.9995 & 5.1961  \\
    \hline 
    $1.88$&1.9063&1.3910&2.9997 & 5.1961  \\
    \hline
    $1.94$&1.9332 &1.3167&2.9999 & 5.1961  \\
    \hline
    $2.00$&1.9513 &1.2542& 2.9999&  5.1961 \\
    \hline
    $3.00$&1.9998 & 0.7458 &$\approx 3$& 5.1962  \\ \hline
    $4.00$&$\approx 2$ & 0.5398 &$\approx 3$& 5.1962  \\\hline
    Schwarzschild&$2$ & no &3& 5.1962  \\
    \hline
    \end{tabular}
    \caption{Numerical results for various parameters of the de Sitter-Schwarzschild black hole with  different ratio $r_g/r_0$. The parameters are the external event horizon $r^+ _h$, internal event horizon $r^- _h$, photon sphere $r_{sp}$ and the apparent radius of the photon sphere $b_{c}$.}
    \label{table1}
\end{table}
Our interest focuses on understanding how the region near a de-SBH appears to a distant observer when different types of emissions occur nearby. To investigate this, we trace the path of light rays (null geodesics) from the observer's viewpoint backward towards the vicinity of the black hole. Throughout this paper, we suppose that the observer is located at $r_{\rm obs}=10^{5}M$ with $\phi =0$. As been noted in Ref.~\cite{Gralla:2019xty}, a crucial factor in understanding the observed appearance is the total change in the azimuthal angle $\phi$ of the orbital plane for these light rays, which depends on the effective impact parameter $b$ of their trajectories. 

\begin{figure}[htbp] 
    \centering
    \begin{subfigure}[b]{0.28\textwidth}
      \includegraphics[width=\textwidth]{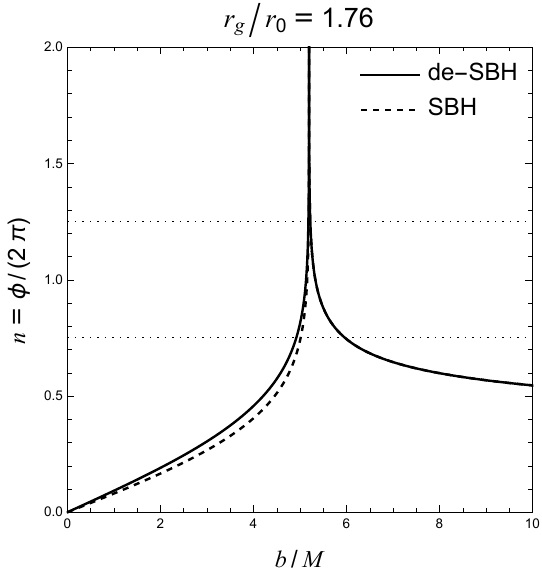}
      \label{fig:numberorbitdSSvsShwarz}
    \end{subfigure} 
    \begin{subfigure}[b]{0.28\textwidth}
      \includegraphics[width=\textwidth]{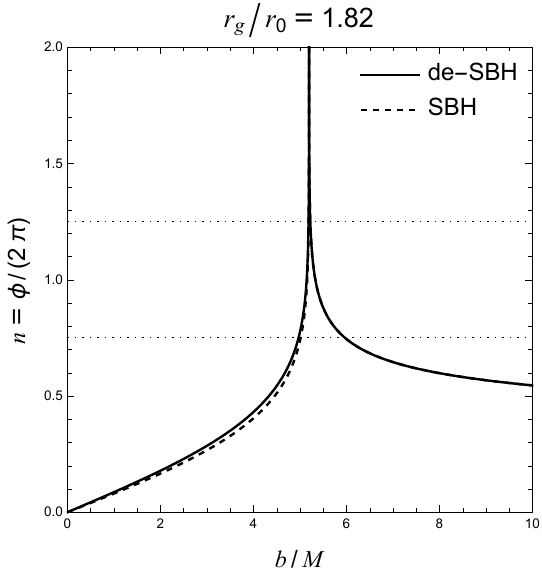}
      \label{fig:numberorbitdSSvsShwarzn182}
    \end{subfigure}
      \begin{subfigure}[b]{0.28\textwidth}
        \includegraphics[width=\textwidth]{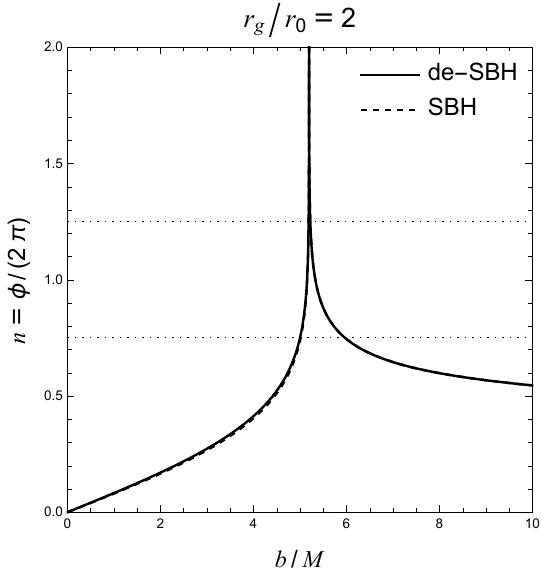}
        \label{fig:numberorbitdSSvsShwarzn2}
      \end{subfigure}
    \caption{Fractional number of photon orbits, represented as $n=\phi/(2\pi)$, for de Sitter-Schwarzschild black holes characterized by different ratios of $r_g/r_0$. For comparison, we also include the corresponding results for the Schwarzschild spacetime, depicted by dashed lines. The dotted lines, $n=0.75$ and $n=1.25$, separate the direct ($n<0.75$), lensed ($0.75<n<1.25$), and photon ring ($n>1.25$) trajectories~\cite{Gralla:2019xty}. }
    \label{fig:numberorbit}
  \end{figure}

In Fig.~\ref{fig:numberorbit}, we show the total number of orbits, $n \equiv \phi/(2 \pi)$, as a function of the effective impact parameter $b$ for de-SBHs with different ratio $r_g/r_0$. For $b<b_c$, the number of orbits for de-SBHs is larger than that for SBH, and this difference diminishes as the ratio $r_g/r_0$ increases.   For $b>b_c$, the photon trajectories are unbounded and stay in the region $r>r_{sp}$ where the difference between the metric of de-SBHs and  that of SBH is negligible (see Fig.~\ref{fig:figmetric} in Appendix~\ref{sec:Reveiw of de Sitter-Schwarzschild spacetime}). Therefore, in the region $r>r_{sp}$, the lensing behavior is indistinguishable between these two kinds of black holes.     

\begin{figure}
    \centering
    \includegraphics[scale=0.85]{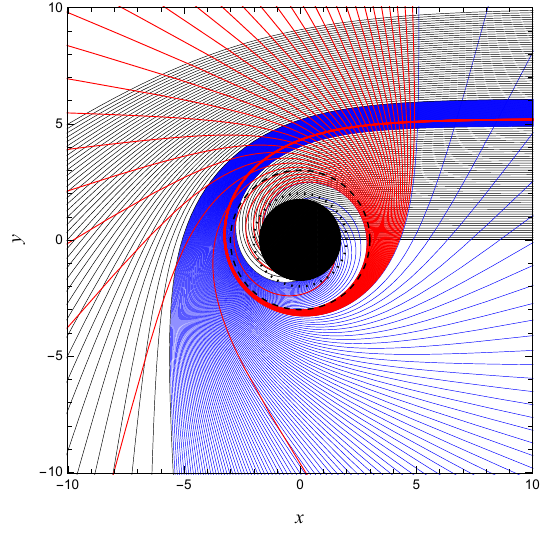}
    \caption{(color online). Photon trajectories in the vicinity of a de Sitter-Schwarzschild black hole with $r_g/r_0=1.76$. The direct ($n<0.75$), lensed ($0.75<n<1.25$) and photon ring ($n>1.25$) trajectories are shown as black, blue and red solid curves, respectively. The de Sitter-Schwarzschild is shown as the solid black disk and the photon sphere is denoted by a black dashed curve. For comparison, the event horizon for the Schwarzschild black hole is depicted by a black dotted curve.}
    \label{fig:dssgeodesic1_distanceobserver}
  \end{figure}

A selection of photon trajectories in the vicinity of a de-SBH with $r_g/r_0=1.76$ is plotted in Fig.~\ref{fig:dssgeodesic1_distanceobserver}, in which the Euclidean coordinates is defined as 
\begin{align}\label{eq:Euclidean coordinates}
    x=r \cos \phi \,,\;\; y=r \sin \phi\,.
\end{align} 

\begin{figure}
    \centering
    \includegraphics[scale=0.3]{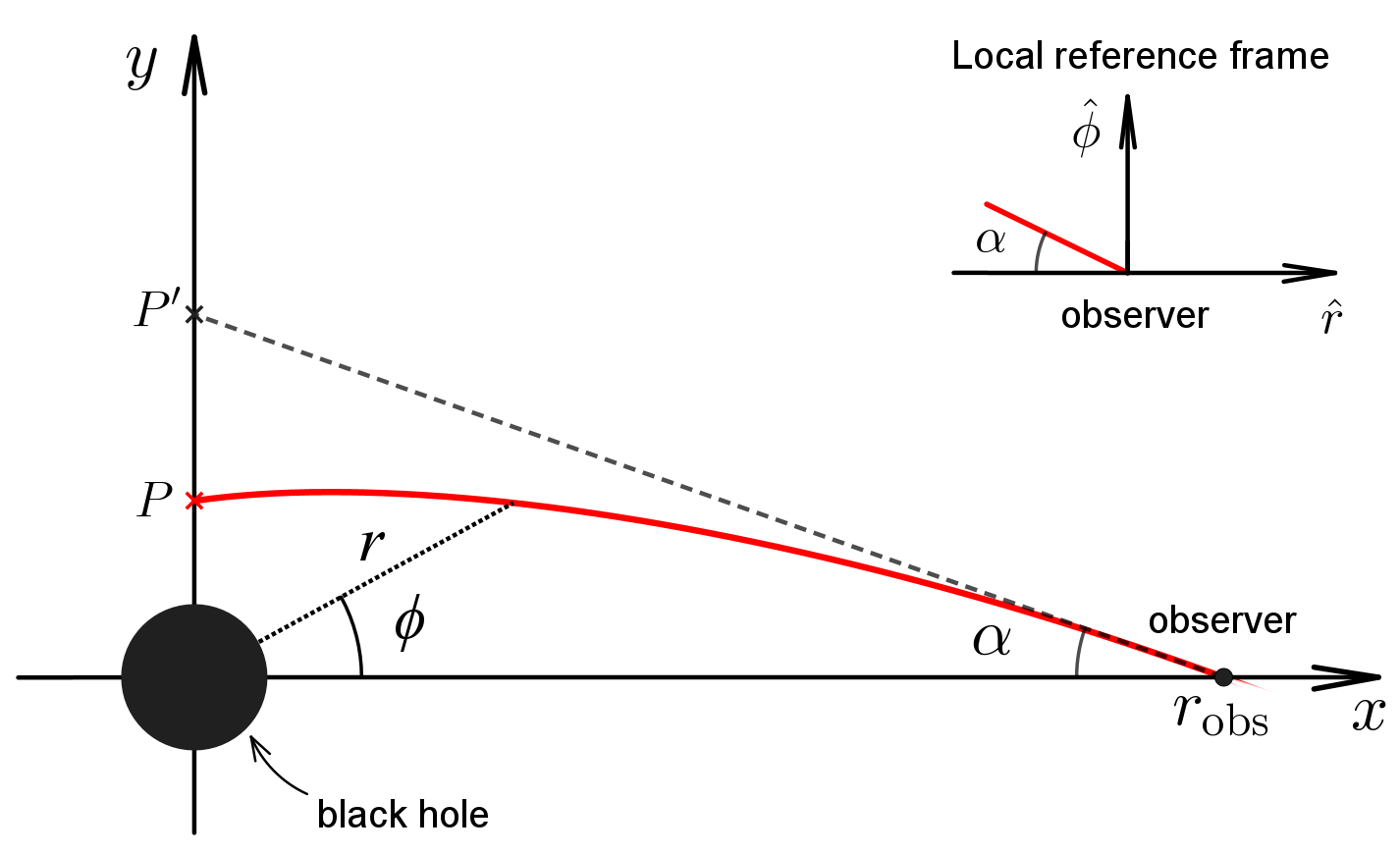}
    \caption{(color online). Coordinate systems for the photon trajectory (red curve) in the $\theta=\pi/2$ plane. The main figure shows the Euclidean coordinates ($x,\,y$) as defined by Eq.~\eqref{eq:Euclidean coordinates}, while the mini-figure illustrates the local reference frame of the observer situated at radius $r_{\rm obs}$. The center of the black hole is located at the origin of the Euclidean coordinates ($x,\,y$). From the observer's perspective, light rays emitted from point $P$ appear to originate from point $P'$.}
    \label{fig:sketchup_viewBH}
  \end{figure}

To understand the observed appearance of black holes, it is convenient to work in the observer's local proper reference frame ($\hat{t}$,\,$\hat{r}$,\,$\hat{\theta}$,\,$\hat{\phi}$), which can be obtained by choosing the non-coordinate orthonormal basis. This orthonormal (dual) basis can be expressed in terms of the coordinate (dual) basis as follows:
\begin{align}\label{eq:tetrads}
  \hat{e}_{i}=\partial /\partial \hat{x}^{\mu}=e_i{} ^{\mu}\, (\partial /\partial x^{\mu}) \;\,,\; \hat{e}^i=\dd \hat{x}^{\mu}=e^i{} _{\mu} \dd x^{\mu}\,.
\end{align} 
The components $e^{i}{}_{\mu}$ forms a $4\times 4$ matrix (with the inverse matrix denoted by $e_i{} ^{\mu}$) satisfying 
\begin{align}
  g_{\mu \nu} =e^{i}{}_{\mu}e^{j}{}_{\nu}\,\eta_{ij}\,,
\end{align}
where $\eta_{ij}$ is the Lorentz metric. The $e^{i}{}_{\mu}$ are known as tetrads, or vielbeins. For the de Sitter-Schwarzschild metric, the tetrads are given by 
\begin{align}
    e^{i}{}_{\mu}= \diag[-\sqrt{g_{tt}},\sqrt{g_{rr}},\sqrt{g_{\theta \theta}},\sqrt{g_{\phi \phi}}]\,.
\end{align}

  Fig.~\ref{fig:sketchup_viewBH} displays a representative trajectory of a photon observed by an observer located at a radius of $r_{\rm obs}$. From the observer's perspective, light rays emitted from point $P$ appear to originate from point $P'$. Therefore, the apparent length from the center of the black hole may be written as 
  \begin{align}
    y_{P'} = r_{\rm obs} \tan \alpha\,,
  \end{align} 
where $\alpha$ is the angular radius. In the local reference frame, we have 
\begin{align}\label{eq:local1}
    \tan \alpha =-\dd \hat {\phi}/ \dd \hat{r}\,,
\end{align}
where $(\dd \hat {\phi},\,\dd \hat{r})$ is the infinitesimal spatial separation of the photon trajectory. Using the relation \eqref{eq:tetrads}, we can obtain
\begin{align}\label{eq:local2}
    \frac{\dd \hat {\phi}}{\dd \hat{r}}=r_{\rm obs}\,\left[1-\frac{r_g}{r_{\rm obs}} \left(1-{\rm{e}} ^{-r_{\rm obs}^3/r_0^2 r_g}\right)\right]^{1/2}\frac{\dd \phi}{\dd r}\bigg| _{r_{\rm obs}}\,.
\end{align}

Combining Eqs.~\eqref{eq:photon_orbit}, \eqref{eq:local1} and \eqref{eq:local2}, one can obtain 
\begin{align}
   \sin ^2 \alpha =\frac{b^2}{r_{\rm obs}^2}\left[1-\frac{r_g}{r_{\rm obs}} \left(1-{\rm{e}} ^{-r_{\rm obs}^3/r_0^2 r_g}\right)\right]\,.
\end{align}
For a distant observer, such as $r_{\rm obs} = 10^5 M \gg M$ as chosen in this paper, we can approximate $\sin \alpha \approx \tan \alpha \approx \alpha$. Then, one can have 
\begin{align}
     y_{P'} \approx b \equiv J/E \,,
\end{align}
where we have assumed $r_0 \lesssim r_{\rm obs}$.  Therefore, $b$ is indeed the impact parameter for null geodesics that reach the observer at infinity. Moreover, in what follows, we will treat $b$ as the apparent length of optical sources from the center of the black hole.

\section{Shadows and rings of the de Sitter-Schwarzschild black hole with thin disk accretion}
\label{sec:Shadows and rings-disk}

In this section, we focus on simple scenarios where the emission originates from an optically and geometrically thin static disk located near the black hole. The disk is observed in a face-on orientation and its specific intensity, denoted as $I_\nu$ (with $\nu$ being the frequency in a static frame), depends solely on the radial coordinate. While considering the face-on disk case, it is important to note, as highlighted in Ref.~\cite{Gralla:2019xty}, that more realistic scenarios involving orbiting and/or infalling matter can be considered. However, for the face-on disk configuration, these effects are degenerate with the choice of the radial profile.

 In general, the (total) invariant intensity, defined by $\mathcal{I}_{\nu} \equiv I_{\nu}/ \nu ^3$, satisfies the relativistic radiative transfer equation~\cite{mihalas1984foundations}:
\begin{align}\label{eq:invariantI}
    \frac{\dd \mathcal{I}_{\nu}}{\dd \lambda}=\frac{j (\nu)}{\nu ^2}- \nu \mathcal{I}_{\nu}  \chi (\nu)\,,
\end{align}
where $j (\nu)$ and $\chi (\nu)$ being emission and absorption coefficient, respectively. Specifically, $j (\nu)$ is emissivity per unit volume in the rest frame of the emitter. For the purpose of this paper, we will neglect absorption, implying that $\chi (\eta)$ is negligible. Additionally, in the scenario discussed in this section, we will introduce a further simplification. In situations where the light ray intersects the accretion disk multiple times, the total specific intensity increases and can be described by 
\begin{align*}
    I^{\rm tot}_{\nu}=I_{\nu}+ I^{\rm add}_{\nu}\,,
\end{align*}
where  $I^{\rm add}_{\nu}$ is the additional intensity resulting from these extra intersections (The intensity due to these extra intersections will be discussed later.) However, the original invariant intensity ${I}_{\nu}/\nu^3$ is still constant along the photon trajectory in our scenario.
Therefore, the emitted specific intensity $I _{\rm em} (r,\nu)$ and the observed specific intensity $I _{\rm obs} (r_{\rm obs},\nu_{\rm obs})$ satisfy the following equation
\begin{align}\label{eq:obs_I}
   \frac{ I _{\rm em} (r,\nu)}{I _{\rm obs} (r_{\rm obs},\nu_{\rm obs})}=\left(\frac{ \nu }{\nu_{\rm obs} }\right)^3 =\left[\frac{g(r_{\rm obs})}{g(r)}\right]^3\,,
\end{align}  
where 
\begin{align}\label{eq:redshift_static}
  g(r)=\left[1-{r_g} \left(1-{\rm{e}} ^{-r^3/r_0^2 r_g}\right)/{r}\right]^{1/2}
\end{align}
is  the redshift factor. For a distant observer, Eq.~\eqref{eq:obs_I} gives 
\begin{align}
    I_{\rm obs} (r_{\rm obs},\nu_{\rm obs})=g^3 I _{\rm em} (r,\nu)\,.
\end{align}
The total observed intensity resulting from light rays emitted from a specific location $r$ is given by 
\begin{align}
    I_{\rm obs} (r_{\rm obs}) = \int I_{\rm obs} (r_{\rm obs},\nu_{\rm obs}) \dd \nu_{\rm obs}=\int  g^4 I _{\rm em} (r,\nu) \dd \nu= g^4 I_{\rm em}(r)\,,
\end{align}
where $I_{\rm em}(r) \equiv \int   I _{\rm em} (r,\nu) \dd \nu $ is the integrated intensity at radial $r$. 

\begin{figure}[tbp] 
    \centering
    \includegraphics[scale=1]{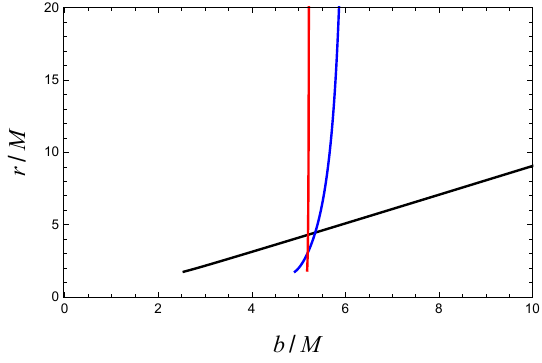}
    \caption{(color online). The first three transfer functions of the de Sitter-Schwarzschild black hole with $r_g/r_0=1.76$. These transfer functions represent the radial coordinates of the first (black), second (blue), and third (red) intersections of a photon traced back from an observer located at $r_{\rm obs}=10^5 M$ with a face-on thin disk outside the event horizon.}
    \label{fig:transferfig}
  \end{figure}

As we mentioned earlier, if a light ray is traced backward from the observer, it may intersect the accretion disk, thereby picking up brightness from the disk emission. The brightness of the observed image depends on the number of times the light ray intersects the accretion disk. Each intersection contributes to the total observed intensity, and the overall brightness is obtained by summing these contributions, as described in Ref.~\cite{Gralla:2019xty},
\begin{align}
    I_{\rm obs} (b)=\sum_{m}\left[g^4 I\right]\Big|_{r=r_m(b)}\,.
\end{align}
Here, $r_m(b)$ ($m=1,2,3,...$) refers to the transfer function, which denotes the radial coordinate of the $m$-th intersection position with the disk plane outside the event horizon. Fig.~\ref{fig:transferfig}  illustrates the first three transfer functions for a de-SBH with $r_g/r_0=1.76$. It is noteworthy that for the de-SBH being studied, the dark region is smaller in comparison to the SBH, despite the photon sphere radii being nearly identical in both cases (see Table~\ref{table1}). Specifically, in the de-SBH, the boundary of the dark region is defined by an impact parameter $b\lesssim 2.6M$, where none of the transfer functions exhibit support. On the other hand, for the SBH, the dark region extends up to $b\lesssim 2.9M$~\cite{Gralla:2019xty}. This variation in the size of the dark region is a consequence of the discrepancy in the radius of the event horizon between these two types of black holes. In the following investigation, we will explore whether this approximately $10\%$ difference in the size of the dark region can manifest as discernible differences in the observational appearance.

\begin{figure}[htbp] 
  \centering \hspace{-8mm}
  \begin{subfigure}[t]{0.28\textwidth}
    \includegraphics[width=\textwidth]{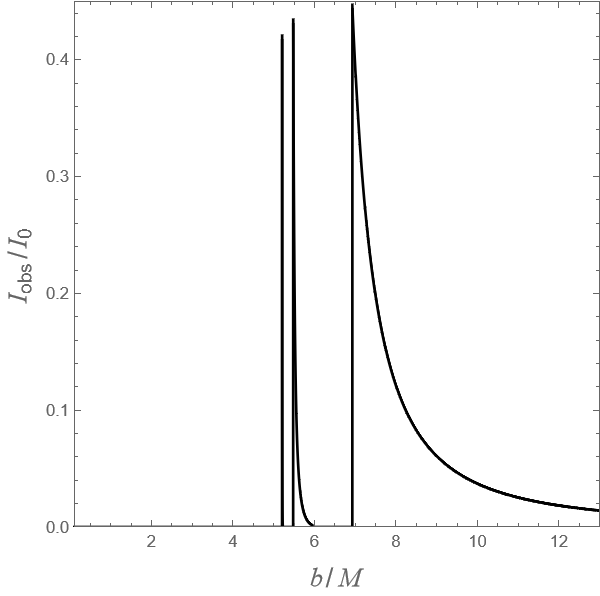}
    \caption*{\quad Emission 1}
    \label{fig:observerI1}
  \end{subfigure} \hspace{8mm}
    \begin{subfigure}[t]{0.28\textwidth}
      \includegraphics[width=\textwidth]{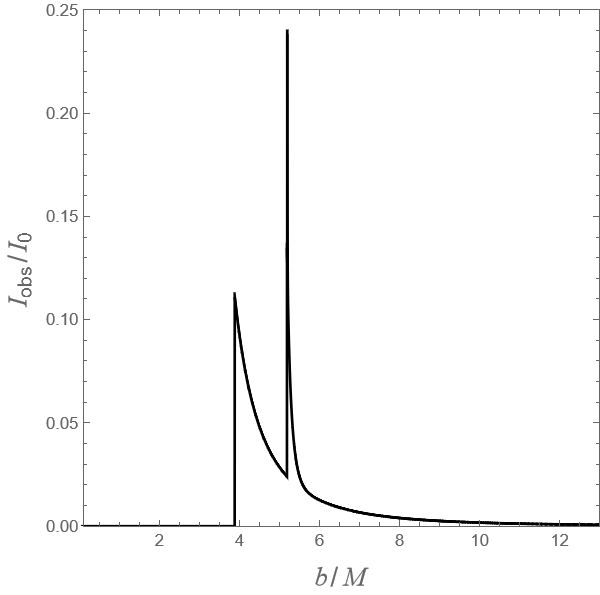}
      \caption*{\quad Emission 2}
      \label{fig:observerI2}
    \end{subfigure}
    \\ \hspace{6mm}
  \begin{subfigure}[b]{0.3\textwidth}
      \raisebox{-0.0\height}{\includegraphics[width=\textwidth]{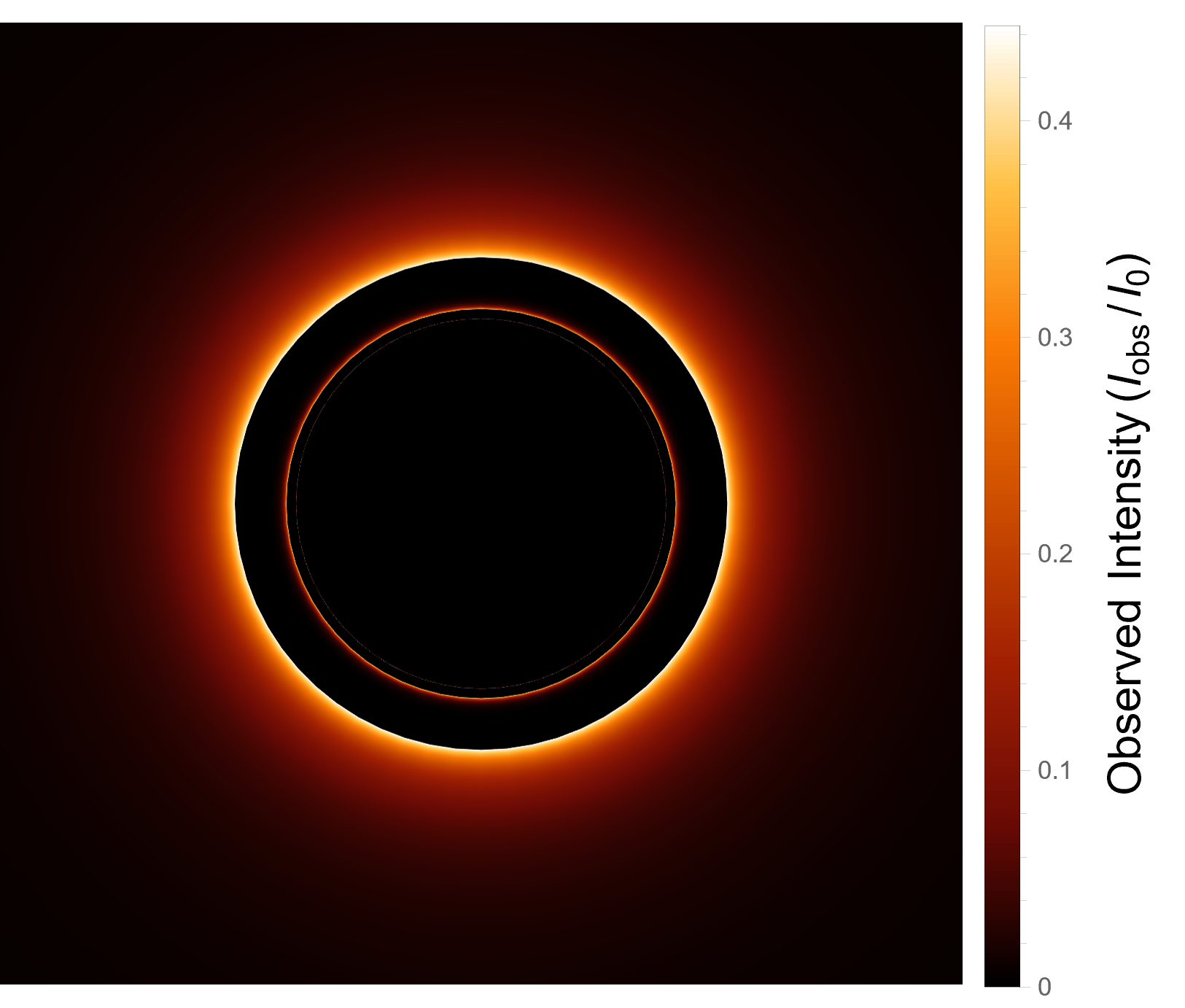}}
      \label{fig:dssBHshadowfinal1}
    \end{subfigure} \hspace{6mm}
    \begin{subfigure}[b]{0.3\textwidth}
        \raisebox{-0.0\height}{\includegraphics[width=\textwidth]{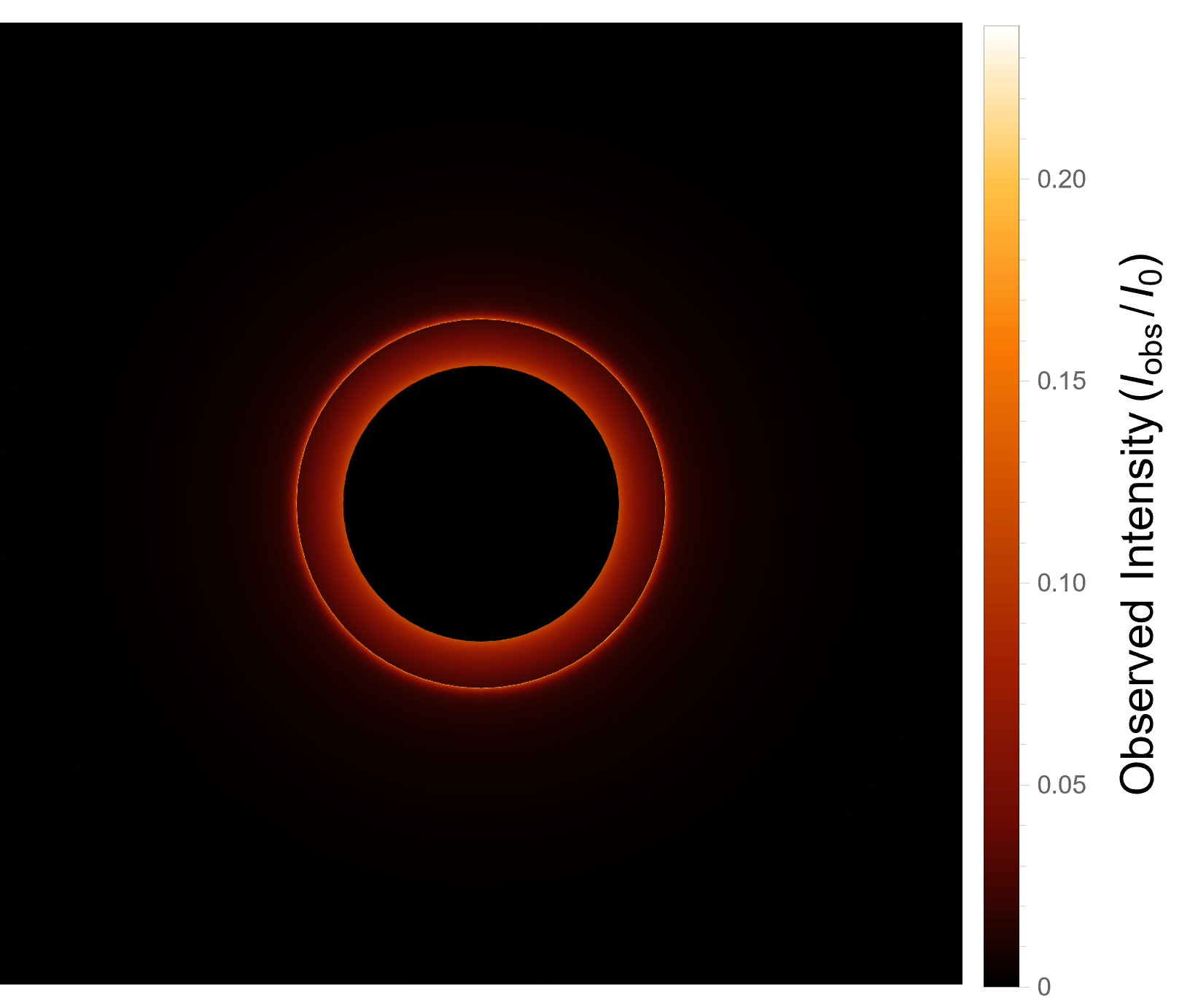}}
        \label{fig:dssBHshadowfinal2}
      \end{subfigure}
  \caption{Observed appearance of a de Sitter-Schwarzschild black hole ($r_g/r_0=1.76$) with an optically and geometrically thin disk accretion, viewed from a face-on orientation. Top panel: Observed intensity received by an observer located at $r_{\rm obs}=10^5M$. Bottom panel: Optical image of the de Sitter-Schwarzschild black hole. The  observed intensity, $I_{\rm obs}$, is normalized to the maximum value $I_0$ of the emitted intensity. The radial profile of emission from left to right is given by Eq.~\eqref{eq:em1} and Eq.~\eqref{eq:em2}, respectively. }
  \label{fig:shadow1}
\end{figure}

We begin with  three kinds of emission intensity profiles which are widely considered in the literature~\cite{Gralla:2019xty,Zeng:2020vsj,Wang:2022yvi,Wang:2023vcv}.  In the first model, we assume that the emission is sharply peaked near $r=6M$ which is approximately the location of  the innermost stable circular orbit (ISCO) (see Appendix~\ref{sec:Geodesics of de-SBH} for the discussion on stable circular orbit of massive particles.) In this situation,  the profile of the emission intensity is modeled as
\begin{align}\label{eq:em1}
    I_{\rm{em}} ^{\rm case1}(r) & = \left\{\begin{array}{ll}
    I_{0}\left(\frac{1}{r-\left(r_{\rm{ISCO}}-1\right)}\right)^{2}, & r\geq r_{\rm{ISCO}} \\
    0, & r < r_{\rm{ISCO}}
    \end{array}\right. \,,
\end{align}
where $I_{0}$ denotes the maximum value of the emitted intensity, and where $r_{\rm{ISCO}}$ represents the radius of the innermost stable circular orbit. 

\begin{figure}[b] 
  \centering \hspace{0mm}
  \begin{subfigure}[t]{0.28\textwidth}
    \includegraphics[width=\textwidth]{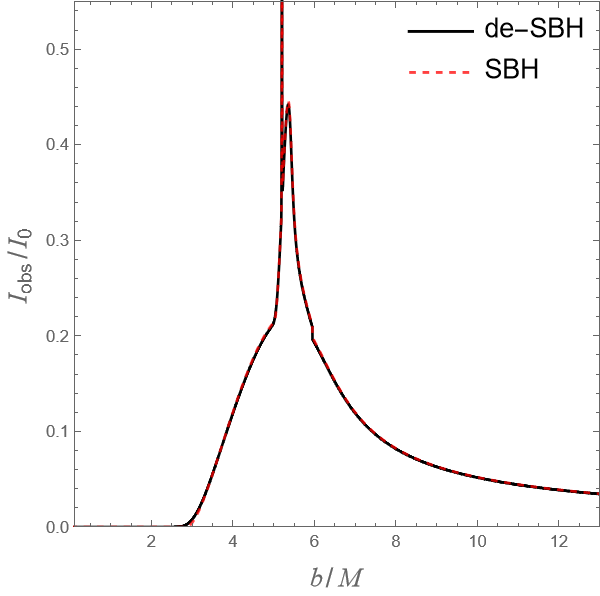}
    \label{fig:compareobserverI3}
  \end{subfigure}\hspace{3mm} 
  \begin{subfigure}[t]{0.28\textwidth}
      \includegraphics[width=\textwidth]{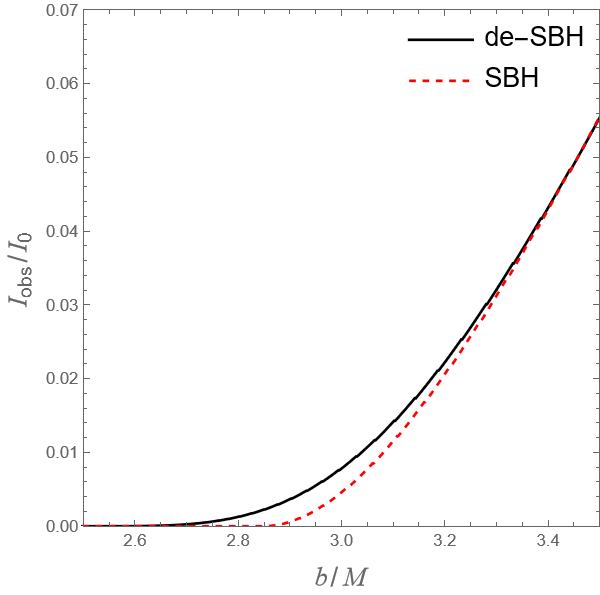}
      \label{fig:compareobserverI3zoom}
    \end{subfigure}    \hspace{3mm}
    \begin{subfigure}[t]{0.3\textwidth}
      \raisebox{0.1\height}{\includegraphics[width=\textwidth]{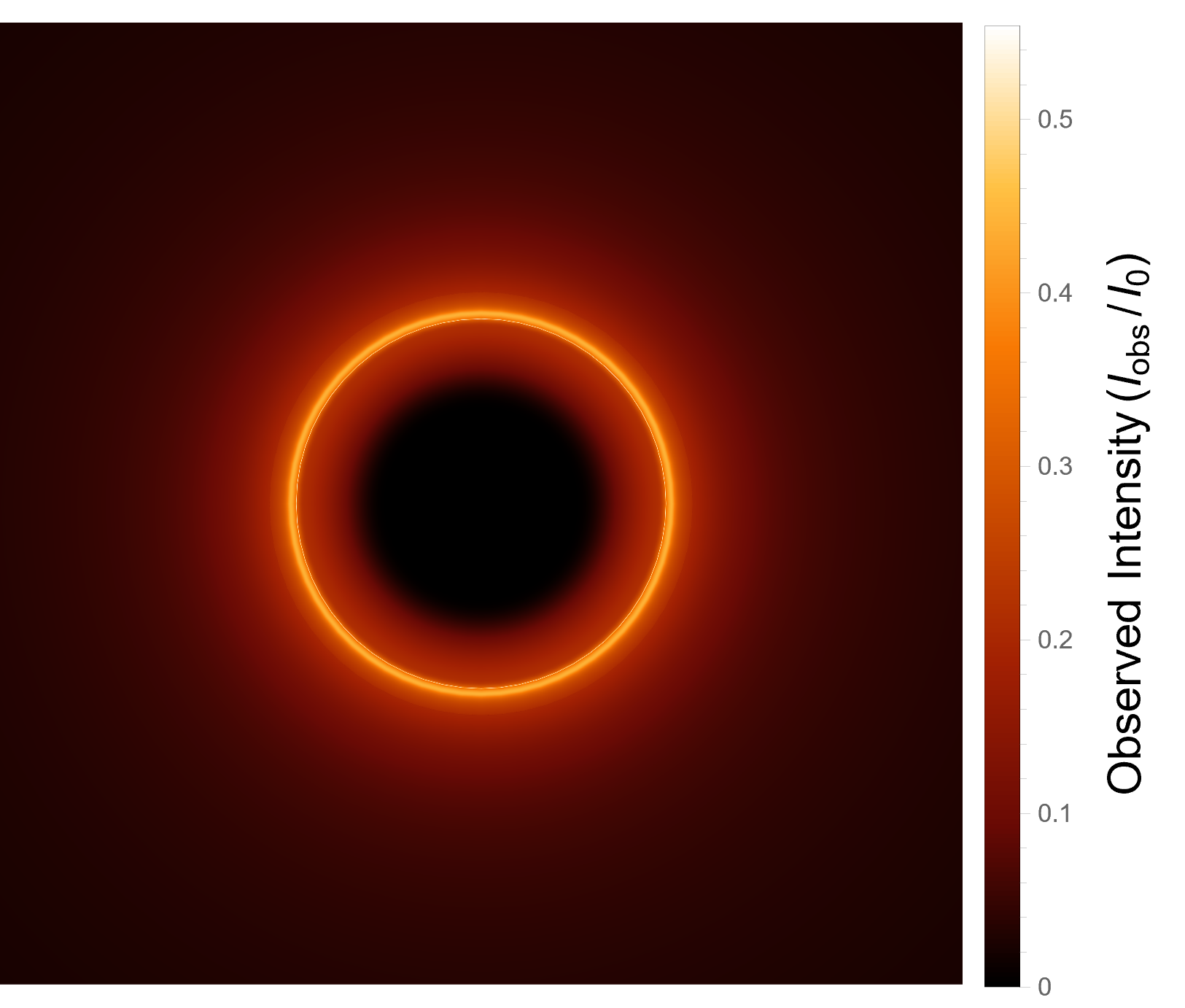}}
      \label{fig:dssBHshadowfinal3}
    \end{subfigure} 
  \caption{Observed appearance of a de Sitter-Schwarzschild black hole ($r_g/r_0 = 1.76$) and
  the Schwarzschild black hole for an optically and geometrically thin disk accretion with emission profile given by Eq.~\eqref{eq:em3}. Left panel: Observed intensity of the de Sitter-Schwarzschild black hole (black solid curve) and Schwarzschild black hole (red dashed curve).  Middle panel: Observed intensity near $b=3M$. Right panel: Optical image of the de Sitter-Schwarzschild black hole.}
  \label{fig:shadow2}
\end{figure}

The second model considers the profile of the emission intensity
\begin{align}\label{eq:em2}
    I_{\rm{em}} ^{\rm case2}(r) & = \left\{\begin{array}{ll}
    I_{0}\left(\frac{1}{r-\left(r_{\rm{sp}}-1\right)}\right)^{3}, & r\geq r_{\rm{sp}} \\
    0, & r < r_{\rm{sp}}
    \end{array}\right. \,,
\end{align}
where the emission is assumed to peak right at the photon sphere, then decay suppressed by the third power. 
For the two types of thin accretion disks mentioned above, we have plotted the observed appearances of a de-SBH with a ratio of $r_g/r_0=1.76$ in Fig.~\ref{fig:shadow1}. It is noteworthy that in these scenarios, both the emission profiles and the observed appearance are practically identical for de-SBHs and SBHs. This finding implies that, based on the considered criteria of emission profiles and observed appearance, it is not possible to distinguish between de-SBHs and Schwarzschild black hole.

\begin{figure}[b] 
  \centering
  \hspace{-8mm}
  \begin{subfigure}[t]{0.28\textwidth}
    \includegraphics[width=\textwidth]{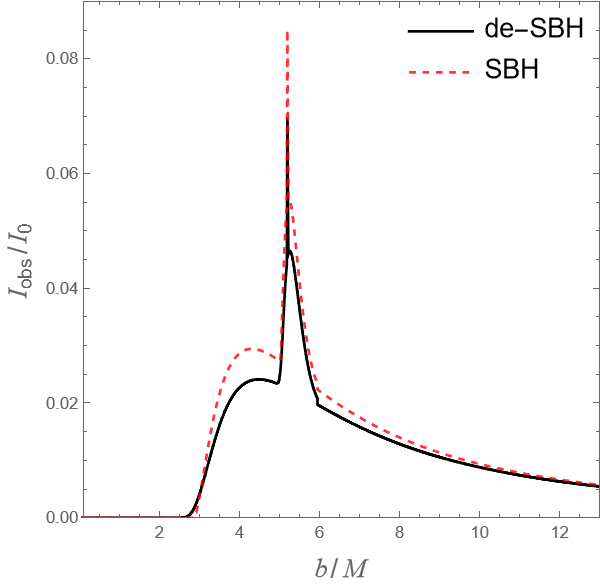}
    \caption*{\quad Emission 4}
    \label{fig:compareobserverI6}
  \end{subfigure}\hspace{8mm}
  \begin{subfigure}[t]{0.28\textwidth}
    \includegraphics[width=\textwidth]{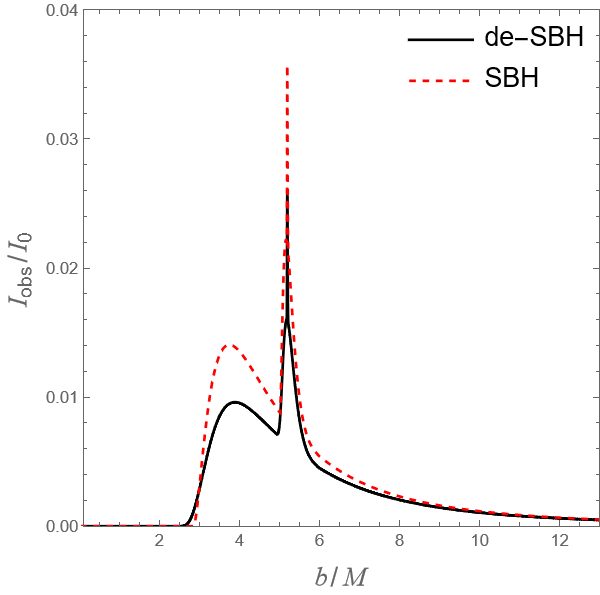}
    \caption*{\qquad Emission 5}
    \label{fig:compareobserverI5}
  \end{subfigure} 
  \\  \hspace{6mm}
  \begin{subfigure}[b]{0.3\textwidth}
      \raisebox{-0.0\height}{\includegraphics[width=\textwidth]{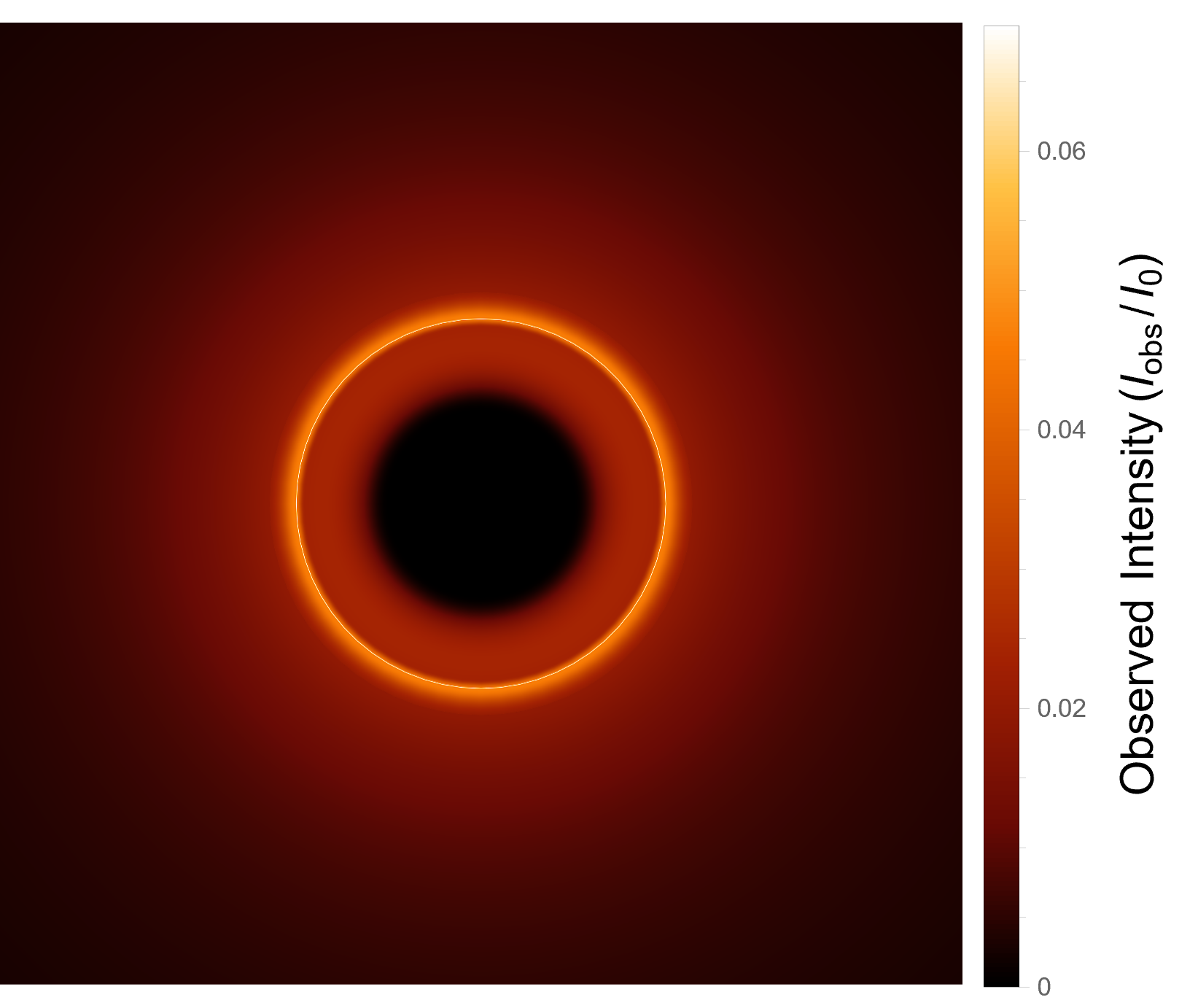}}
      \label{fig:dssBHshadowfinal6}
    \end{subfigure}
\hspace{6mm}
    \begin{subfigure}[b]{0.3\textwidth}
        \raisebox{-0.0\height}{\includegraphics[width=\textwidth]{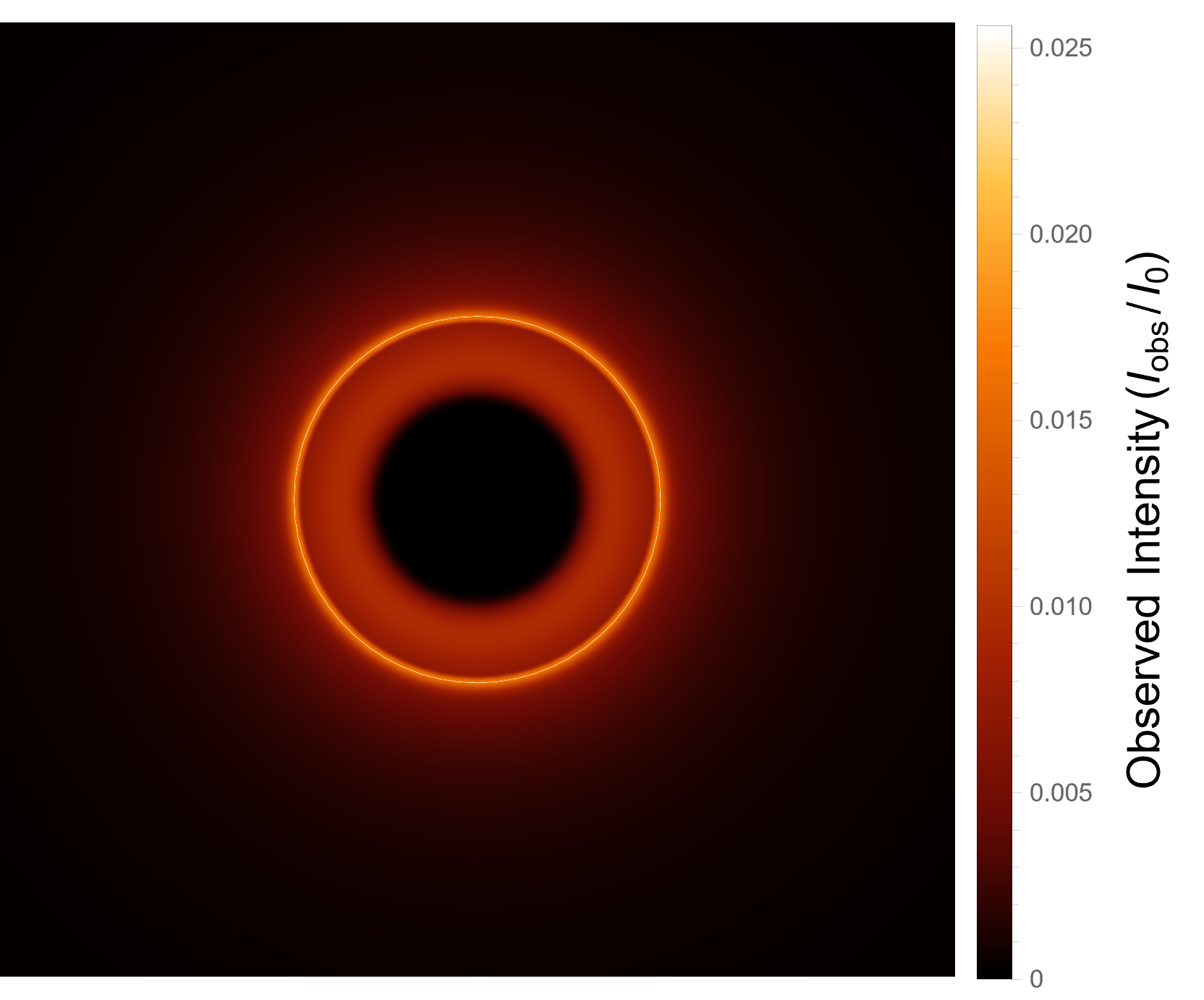}}
        \label{fig:dssBHshadowfinal5}
      \end{subfigure}
  \caption{Observed appearance of a de Sitter-Schwarzschild black hole ($r_g/r_0 = 1.76$) and
  the Schwarzschild black hole with an optically and geometrically thin disk accretion.  Top panel: Observed intensity of the de Sitter-Schwarzschild black hole (black solid curve) and Schwarzschild black hole (red dashed curve). Bottom panel: Optical image of the de Sitter-Schwarzschild black hole. The radial profile of emission from left to right is given by Eq.~\eqref{eq:em4} and Eq.~\eqref{eq:em5}, respectively.}
  \label{fig:shadow3}
\end{figure}

\begin{figure}
  \begin{subfigure}[t]{0.28\textwidth}
    \raisebox{-0.1\height} {\includegraphics[width=\textwidth]{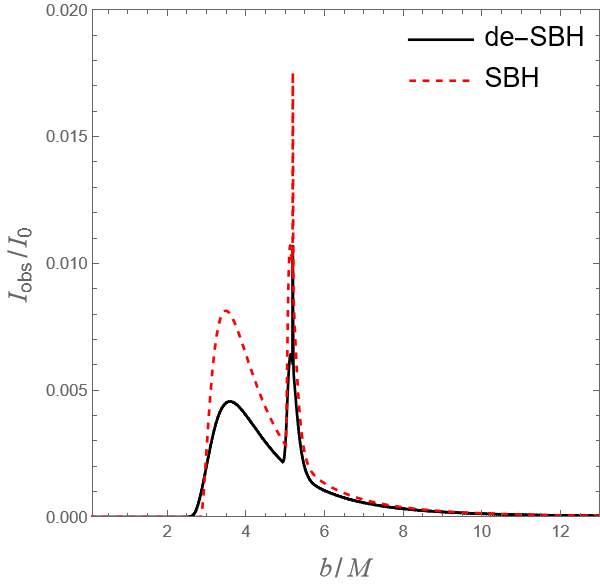}}
    \label{fig:compareobserverI4}
  \end{subfigure} \hspace{4mm}
\begin{subfigure}[t]{0.3\textwidth}
  \includegraphics[width=\textwidth]{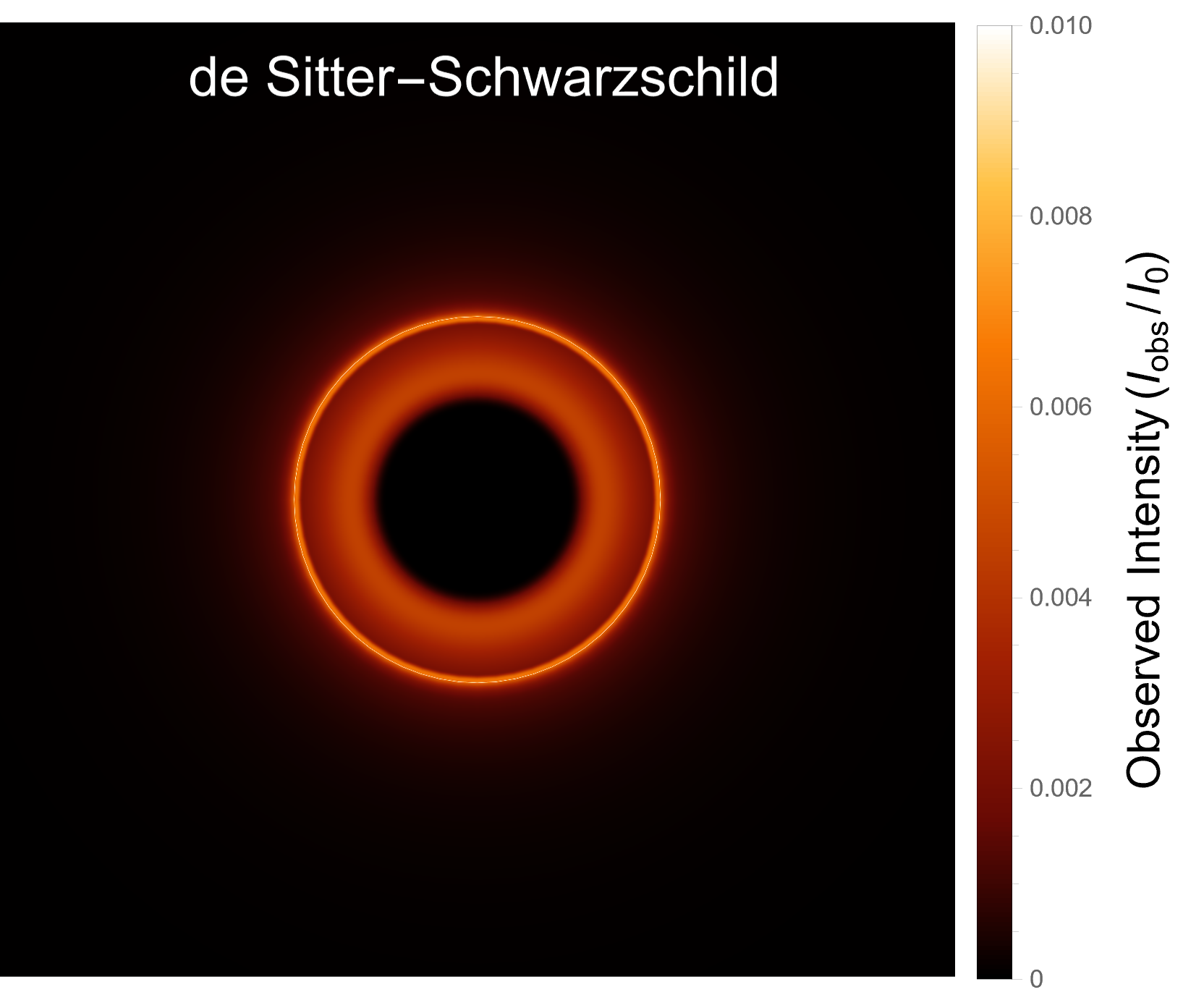}
  \label{fig:dssBHshadowfinal4a}
\end{subfigure} \hspace{2mm}
\begin{subfigure}[t]{0.3\textwidth}
  \includegraphics[width=\textwidth]{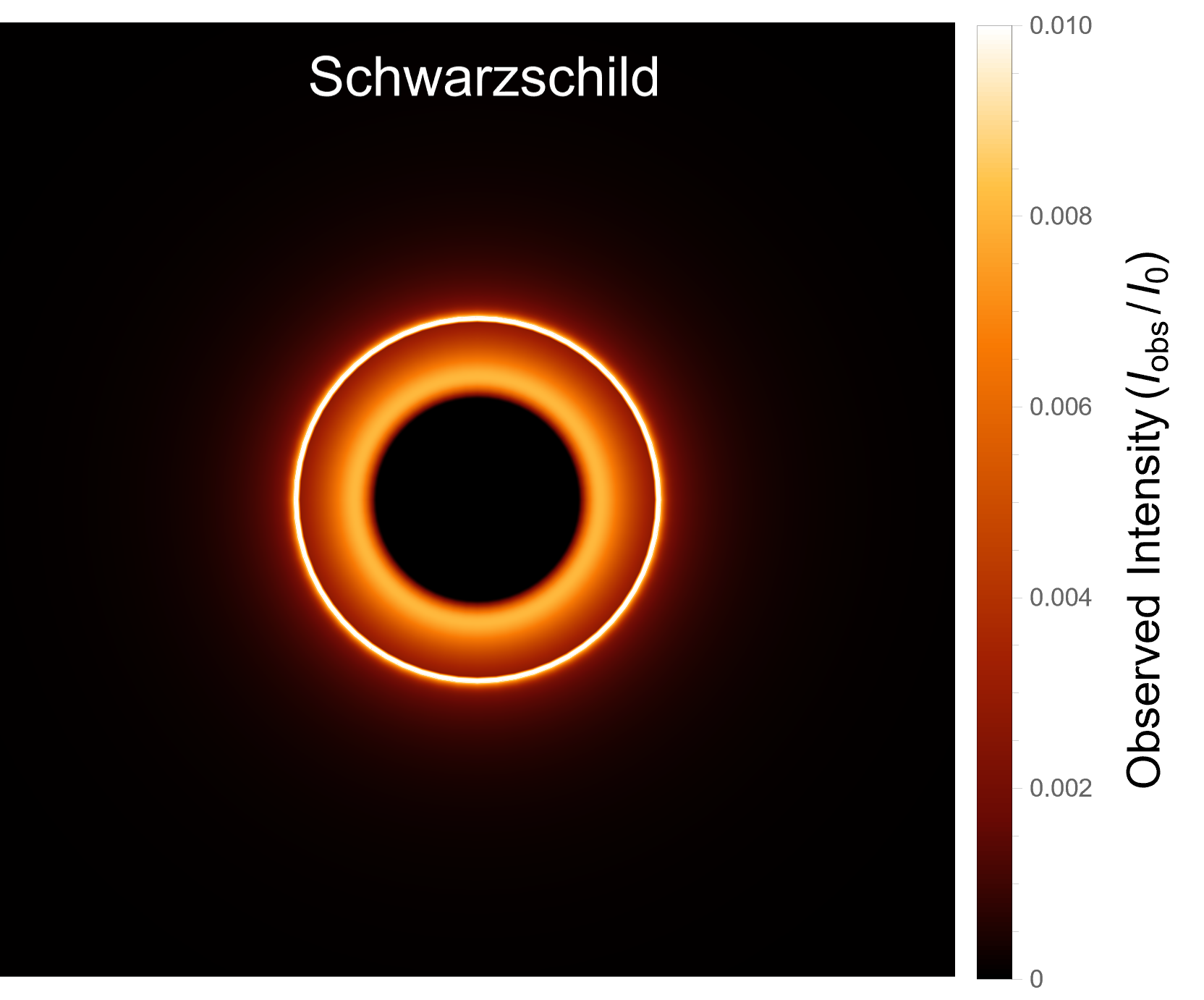}
  \label{fig:dssBHshadowfinalSbh4}
\end{subfigure}
\caption{Observed appearance of a de Sitter-Schwarzschild black hole ($r_g/r_0=1.76$) for an optically and geometrically thin disk accretion with emission profile given by Eq.~\eqref{eq:em6}. Left panel: Observed intensity of the de Sitter-Schwarzschild black hole (black solid curve) and Schwarzschild black hole (red dashed curve).  Middle panel: Optical image of the de Sitter-Schwarzschild black hole. Right panel: Optical image of the Schwarzschild black hole.}
\label{fig:shadow4}
\end{figure}

In the third model,  the emission profile starts from the event horizon and exhibits a relatively slow decay as it approaches the innermost stable circular orbit. The expression for this emission profile is modeled as follows:
\begin{align}\label{eq:em3}
    I_{\rm{em}} ^{\rm case3}(r)=\left\{\begin{array}{ll}
        I_{0} \frac{\frac{\pi}{2}-\arctan \left[r-\left(r_{\rm{ISCO}}-1\right)\right]}{\frac{\pi}{2}-\arctan \left[r_{h}-\left(r_{\rm{ISCO}}-1\right)\right]}, & r \geq r_{h} \\
        0, & r < r_{h}
        \end{array}\right.\,.
\end{align}
The observed appearance of the de-SBH and its comparison with the SBH is depicted in Fig.~\ref{fig:shadow2}. Upon analyzing the middle panel of Fig.~\ref{fig:shadow2}, we can observe a subtle difference in the observed intensities between these two types of black holes, particularly around $b\sim 3 M$, where the de-SBH exhibits slightly higher brightness. However, this difference is relatively small when considering the overall brightness, as illustrated in the left panel of Fig.~\ref{fig:shadow2}. Consequently, the observed appearances of these black holes appear virtually indistinguishable to the naked eye.

In order to simulate the emission from matter falling into the black hole, we now explore three additional types of emission intensity profiles characterized by sharp peaks near the event horizon, followed by different decay rates 
  \begin{subequations}
    \begin{align}\label{eq:em4}
      I_{\rm{em}} ^{\rm case4}(r) & = \left\{\begin{array}{ll}
      I_{0}\left(\frac{1}{r-\left(r_{{h}}-1\right)}\right)^{2}, & r\geq r_{{h}} \\
      0, & r < r_{{h}}
      \end{array}\right. \,,\\ \label{eq:em5}
    I_{\rm{em}} ^{\rm case5}(r) & = \left\{\begin{array}{ll}
    I_{0}\left(\frac{1}{r-\left(r_{{h}}-1\right)}\right)^{3}, & r\geq r_{{h}} \\
    0, & r < r_{{h}}
    \end{array}\right. \,,\\ \label{eq:em6}
    I_{\rm{em}} ^{\rm case6}(r) & = \left\{\begin{array}{ll}
    I_{0}\left(\frac{1}{r-\left(r_{{h}}-1\right)}\right)^{4}, & r\geq r_{{h}} \\
    0, & r < r_{{h}}
    \end{array}\right. \,.
  \end{align} 
\end{subequations}
In Figs.~\ref{fig:shadow3} and~\ref{fig:shadow4}, we present the observed appearance of the de-SBH for these emission profiles. Notably, despite having similar slopes in the observed intensity for both the de SBH and SBH (assuming the same emission profile), their magnitudes exhibit obvious differences. This distinction is readily apparent from the top panel of Fig.~\ref{fig:shadow3} (see also Fig.~\ref{fig:shadow4}). 

\section{Shadows and rings of the de Sitter-Schwarzschild black hole with thin spherical emission}
\label{Optically and geometrically thin spherical emission}

Next, we investigate the shadows and rings of the de-SBH with static and infalling spherical emission. According to Eq.~\eqref{eq:invariantI}, the observed specific intensity can be obtained by integrating the specific emissivity along the light ray 
\begin{align}\label{eq:sphere accretion0}
  I_{\rm obs} (\nu _{\rm obs}, r_{\rm obs})=\nu _{\rm obs} ^3 \int _{\rm ray} \frac{j(\nu _{\rm em})}{\nu _{\rm em}^2} \dd \lambda\,,
\end{align}
where the absorption is neglected.

Note that the photon frequency measured by an observer with 4-velocity $u^{\alpha}$ is given by $\nu=-g_{\alpha \beta}k^{\beta}u^{\alpha}$, where $k^{\alpha}$ is the 4-momentum of the photon. Therefore, the redshift factor is given by 
\begin{align}\label{eq:sphere accretion1}
  g= \nu_{\rm obs}/\nu_{\rm em}= \frac{k_{\alpha} u^{\alpha}_{\rm obs}}{k_{\beta} u^{\beta}_{\rm em}}\,.
\end{align}
Here, $u^{\beta}_{\rm em}$ corresponds to the 4-velocity of the emitter. We will consider several simple models of specific emissivity with monochromatic emissions at a rest-frame frequency $\nu_\star$,\footnote{A specific rest-frame frequency $\nu_\star$ can be uniquely associated with a particular value of $E$.} characterized by different radial profiles $j_{l}(r)$:
\begin{align}\label{eq:sphere accretion2}
  j(\nu_{\rm em}) = {\delta (\nu_{\rm em}-\nu_{\star})}j_{l}(r)\,.
\end{align} 
The radial profile is modeled as $j_{l} = 1/r^{2l}$, where $l$ takes values of 1, 2, 3, or 4. It is worth noting that the profile $j_1(r) = 1/r^2$ was originally considered in Ref.~\cite{Bambi:2013nla}, while the models with $j \propto 1/r^6$ and $1/r^8$ were discussed in Ref.~\cite{Bauer:2021atk}. Additionally, the profile $j\propto 1/r^4$ is chosen to simulate the model presented in Ref.~\cite{Kocherlakota:2022jnz} (specifically, emission 1 of that reference.)


With the help of Eqs.~\eqref{eq:sphere accretion1} and \eqref{eq:sphere accretion2}, the total observed intensity can be obtained by integrating Eq.~\eqref{eq:sphere accretion0} over all observed frequencies 
\begin{align}
  I_{\rm obs}=-\int _{\rm ray} \frac{j_l (r)\,g^3 k_{\alpha} u^{\alpha}_{\rm obs}}{k^r}\dd r\,,
\end{align}
where the definition  $k^r \equiv  \dd  r/ \dd \lambda$ is used.
For a stationary distant observer with
$
  u^{\alpha}_{\rm obs}= \left\{1,0,0,0\right\}\,,
$
we have 
\begin{align}
  I_{\rm obs}=-\int _{\rm ray} \frac{j_l (r)\,g^3 k_{t} }{k^r}\dd r\,.
\end{align}
The 4-momentum of the photon is given by
\begin{align}
  k_{\alpha}=\left\{-E,\, \pm E \sqrt{1/g_{tt}^2+b^2/(r^2 g_{tt})} ,0,\,bE  \right\}\,,
\end{align}
where the positive ($+$) or negative ($-$) sign corresponds to photons moving away from or approaching towards the black hole, respectively.

Next, we first consider the case of static spherical emissions and subsequently investigate a scenario involving free infalling spherical emissions.

\subsection{Static spherical emissions}
\begin{figure}[htbp] 
  \centering
  \begin{subfigure}[t]{0.3\textwidth}
    \includegraphics[width=\textwidth]{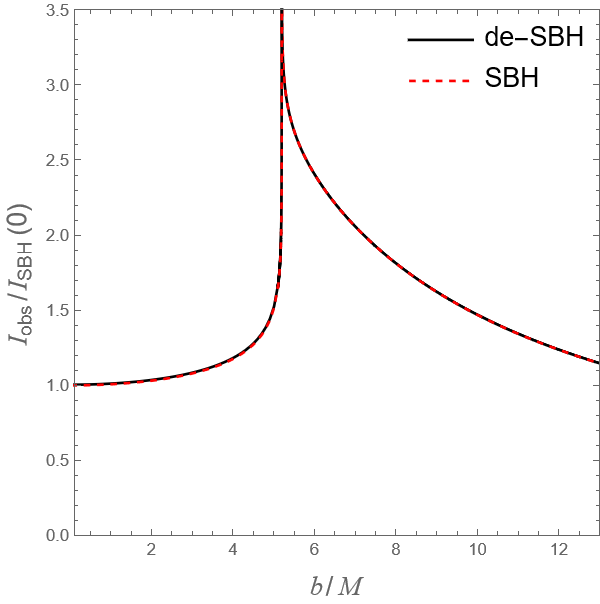}
   \caption*{$j_{1}=1/r^2$}
    \label{fig:datashpereacc1}
  \end{subfigure} \hspace*{2.0mm}
  \begin{subfigure}[t]{0.3\textwidth}
    \includegraphics[width=\textwidth]{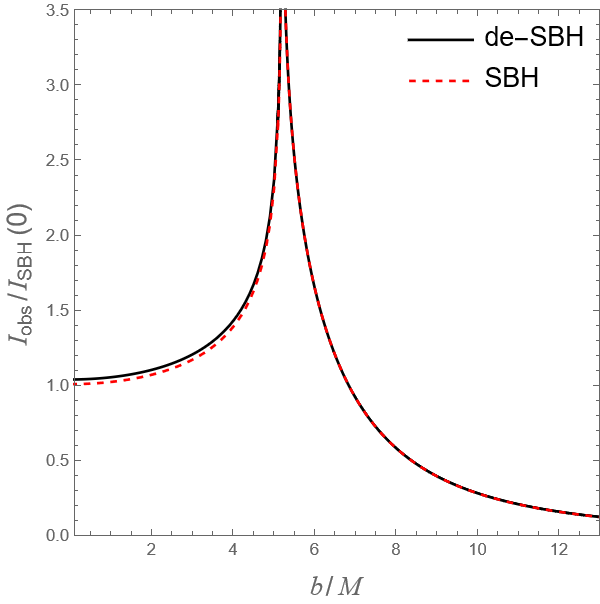}
    \caption*{$j_{2}=1/r^4$}
    \label{fig:datashpereacc2}
  \end{subfigure} \hspace*{2.0mm}
  \begin{subfigure}[t]{0.3\textwidth}
    \includegraphics[width=\textwidth]{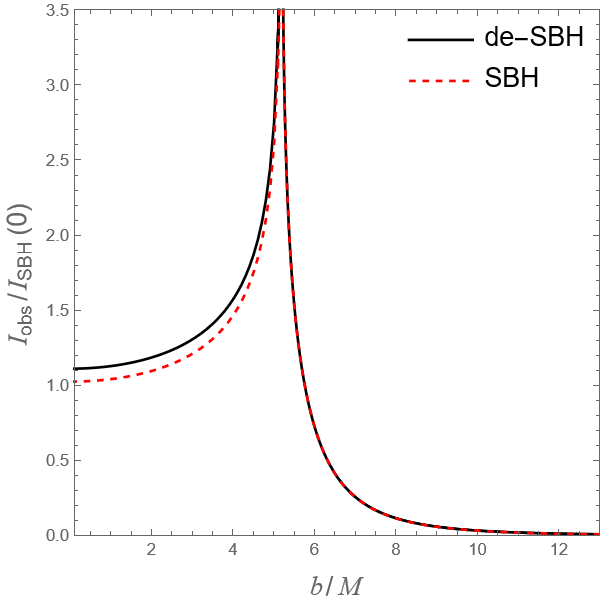}
    \caption*{$j_{3}=1/r^6$}
    \label{fig:datashpereacc3}
  \end{subfigure} \\ \hspace*{6.0mm}
  \begin{subfigure}[t]{0.3\textwidth} 
    \includegraphics[width=\textwidth]{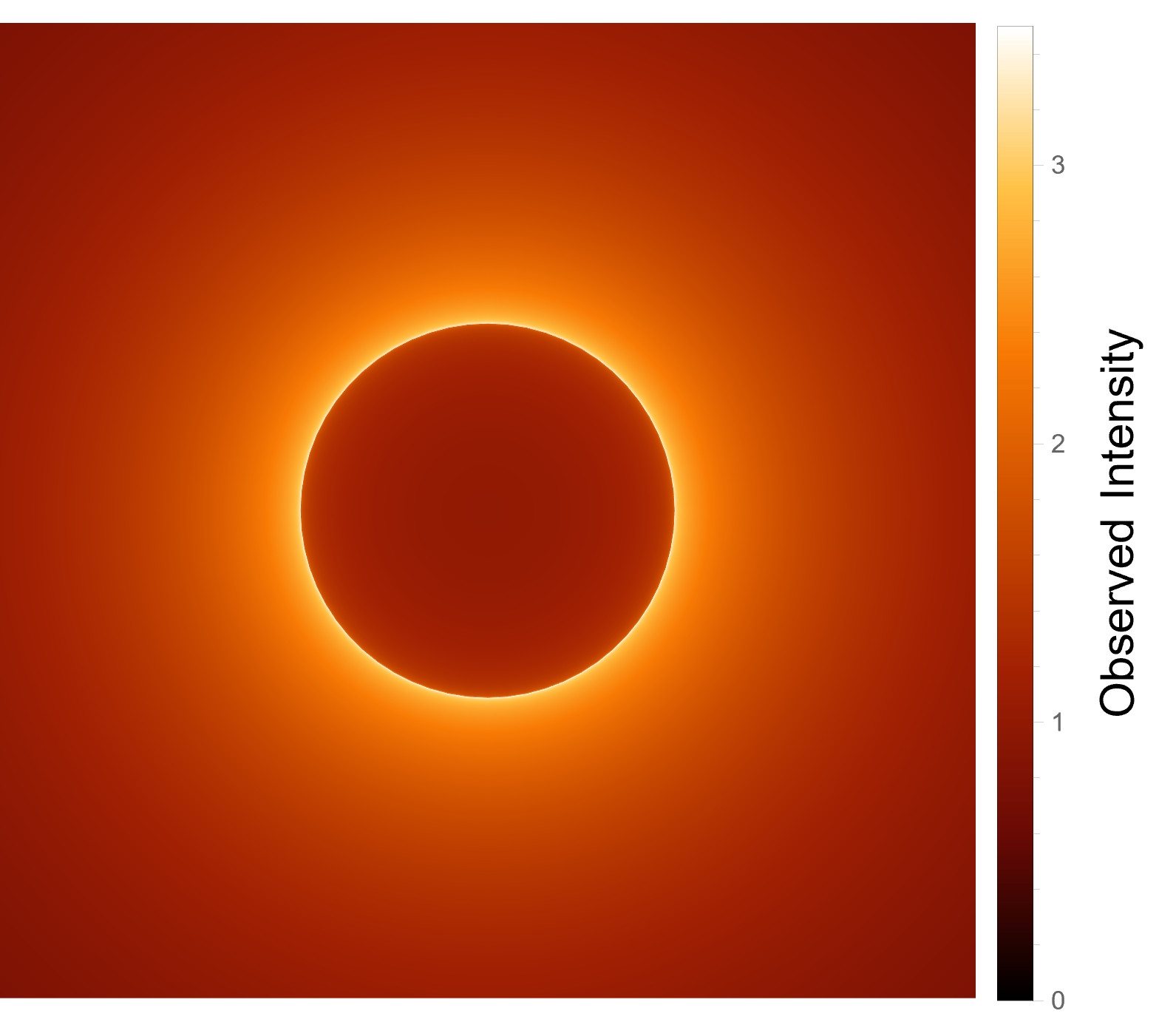}
    \label{fig:dssBHshadowsphere1}
  \end{subfigure} \hspace*{2.0mm}
  \begin{subfigure}[t]{0.3\textwidth} 
    \raisebox{0.0\height}{\includegraphics[width=\textwidth]{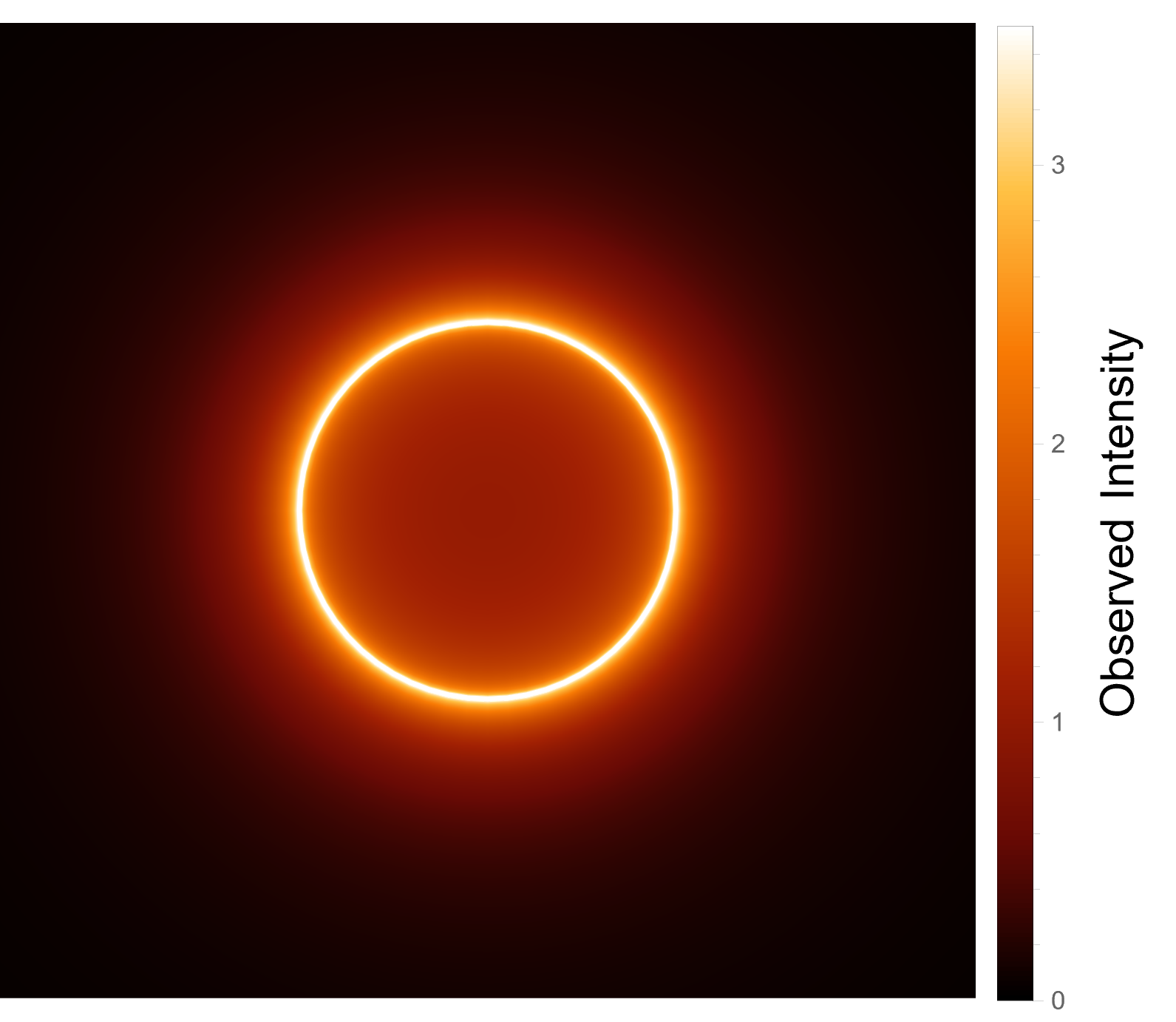}}
    \label{fig:dssBHshadowsphere2}
  \end{subfigure}\hspace*{2.0mm}
  \begin{subfigure}[t]{0.3\textwidth} 
    \raisebox{0.0\height}{\includegraphics[width=\textwidth]{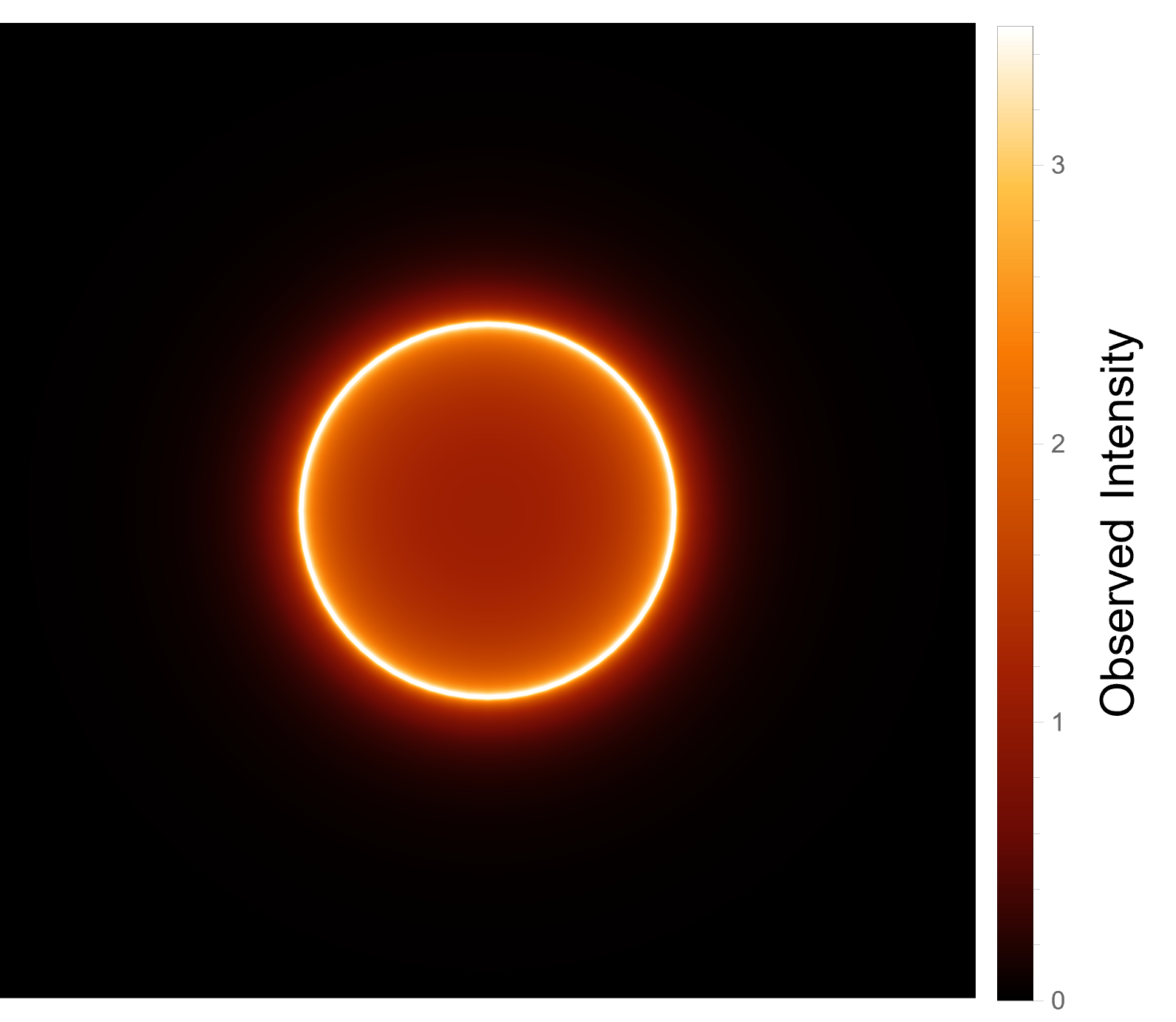}}
    \label{fig:dssBHshadowsphere3}
  \end{subfigure}
  \caption{Observed appearance of a de Sitter-Schwarzschild black hole ($r_g/r_0=1.76$) and the Schwarzschild black hole  with static spherical accretions. Top panel: Observed intensity $I_{\rm obs}/I_{\rm SBH}(0)$ as a function of the impact parameter $b$, where $I_{\rm SBH}(0)$ indicates the observed intensity of the Schwarzschild black hole at $b=0$. Bottom panel: Optical image of the de Sitter-Schwarzschild black hole. The radial profile of emission from left to right is $j_{1}=1/r^2$, $j_{2}=1/r^4$ and $j_{3}=1/r^6$, respectively.   }
  \label{fig:shadow5}
\end{figure}
\begin{figure}[htbp] 
  \centering
  \begin{subfigure}[t]{0.28\textwidth}
    \includegraphics[width=\textwidth]{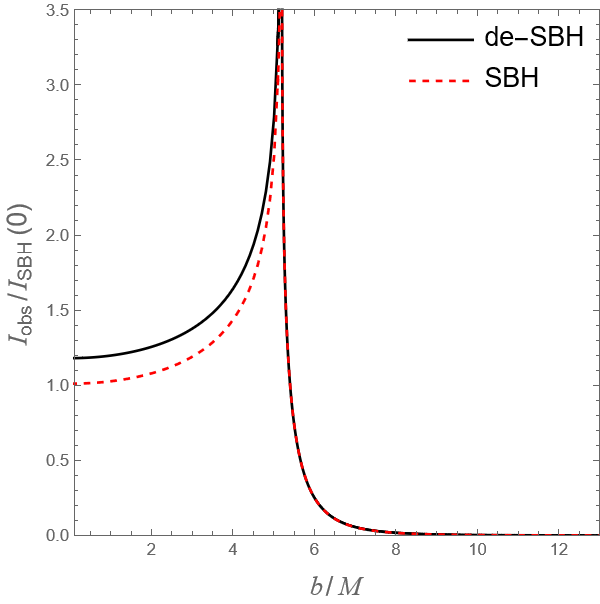}
    \label{fig:datashpereacc4}
  \end{subfigure} \hspace*{3.0mm}
  \begin{subfigure}[t]{0.32\textwidth} 
    \raisebox{0.06\height}{\includegraphics[width=\textwidth]{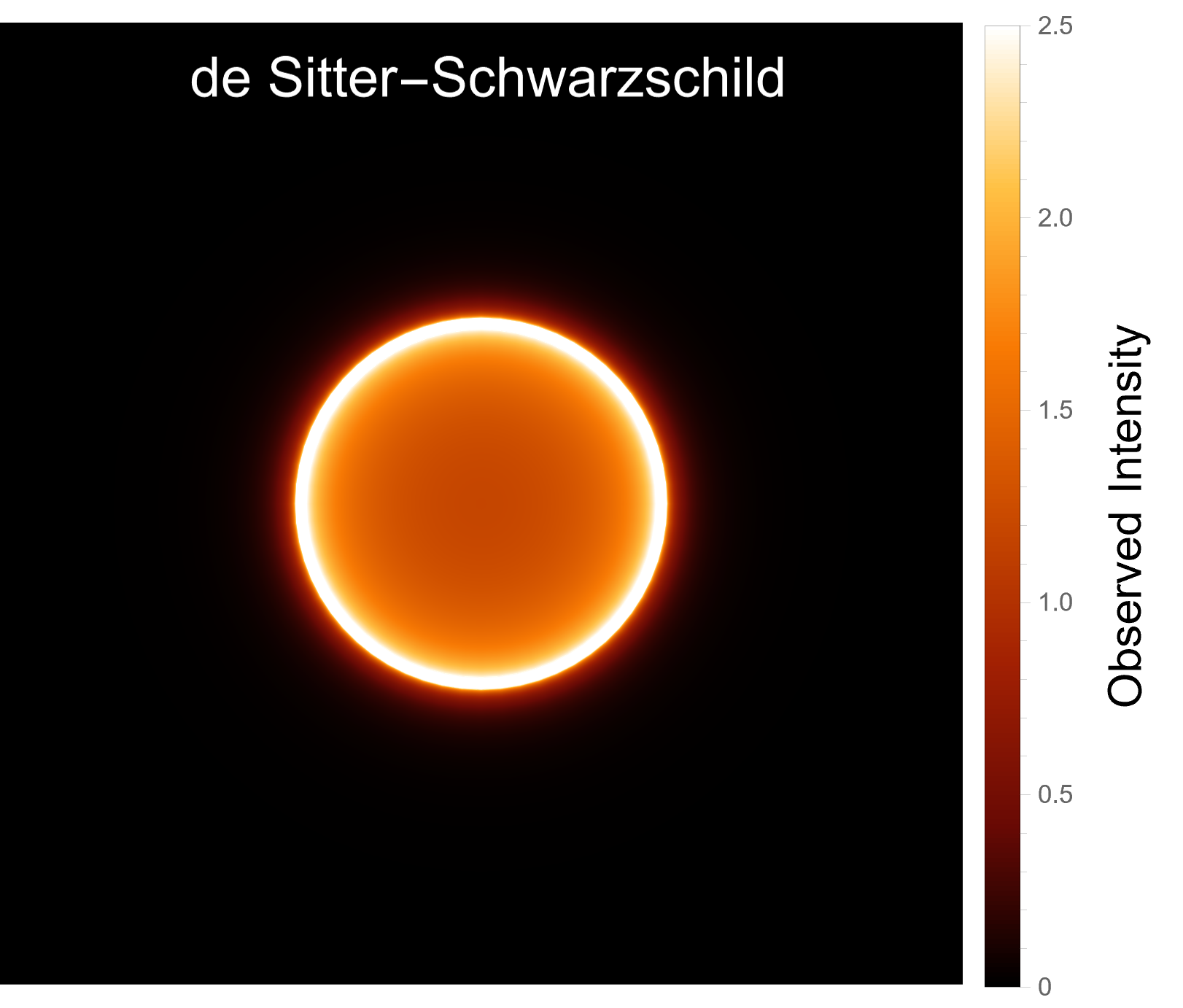}}
    \label{fig:dssBHshadowsphere4}
  \end{subfigure}\hspace*{3.0mm}
  \begin{subfigure}[t]{0.32\textwidth} 
    \raisebox{0.06\height}{\includegraphics[width=\textwidth]{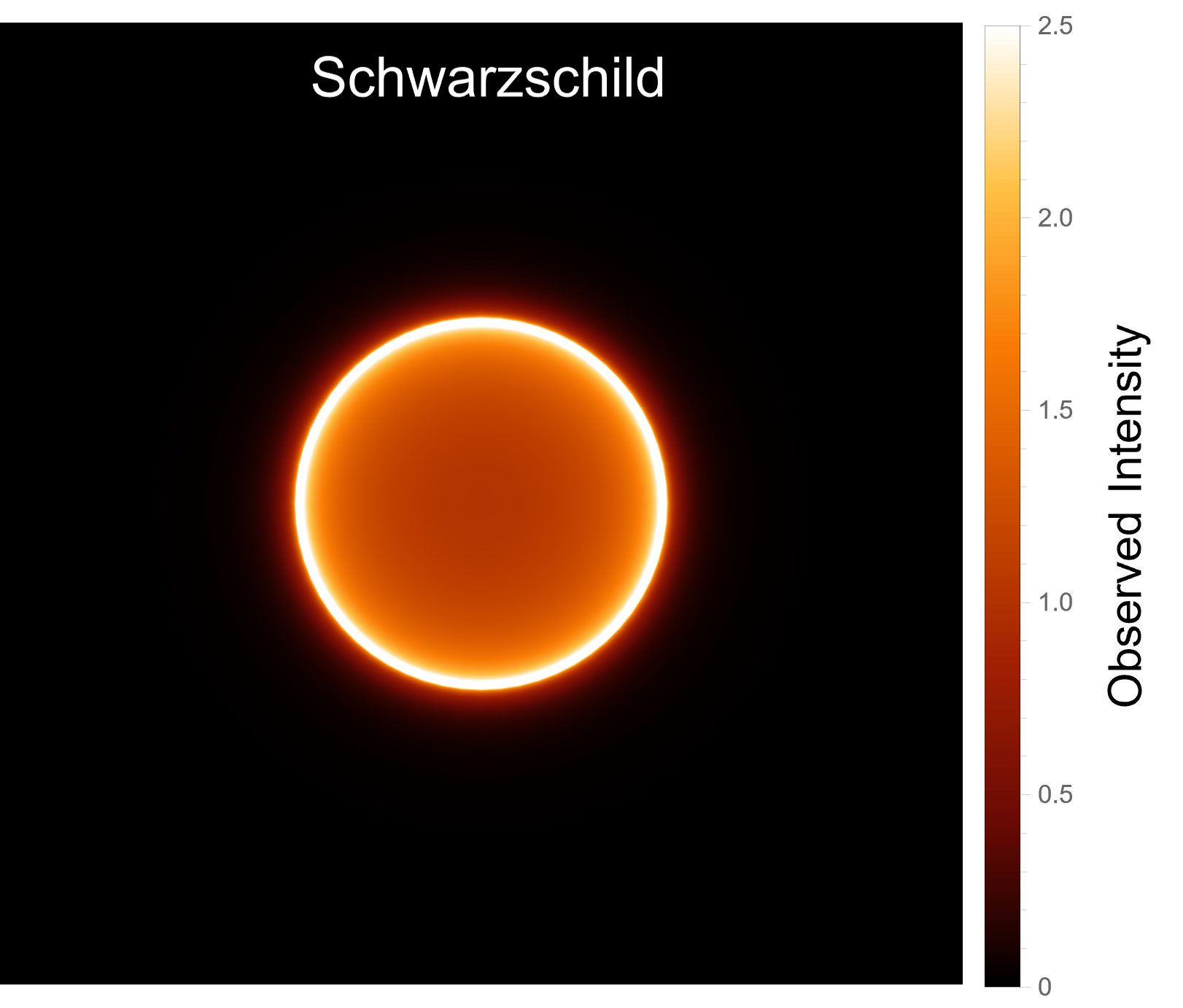}}
    \label{fig:SBHshadowsphere4b.PNG}
  \end{subfigure}
  \caption{Observed appearance of a de Sitter-Schwarzschild black hole ($r_g/r_0=1.76$) and the Schwarzschild black hole  with a static spherical emission profile $j_{4}=1/r^8$. Left panel: Observed intensity $I_{\rm obs}/I_{\rm SBH}(0)$ as a function of the impact parameter $b$. Middle panel: Optical image of the de Sitter-Schwarzschild black hole. Right panel: Optical image of the Schwarzschild black hole.  }
  \label{fig:shadow6}
\end{figure}
For static spherical emissions, we have 
\begin{align}
  u^{\alpha}_{\rm em}=\left(1/\sqrt{-g_{tt}},0,0,0\right)\,,
\end{align}
and the redshift factor $g = \sqrt{-g_{tt}}$ is given by Eq.~\eqref{eq:redshift_static}.
The total observed intensity for static emissions is governed by
\begin{align}\label{eq:Istatic0}
  I_{\rm obs}^{\rm static}=\pm \int _{\rm ray} \frac{r\, g^3 \,j_l}{\sqrt{r^2-b^2g^2}}\dd r\,.
\end{align}
Note that for a light ray that is not captured by a black hole, its trajectory exhibits a ``turning point" at radius $r_{\rm turn}$. This turning point satisfies the equation $r_{\rm turn}^2 = b^2 \,[g(r_{\rm turn})]^2$.  In the case of these light rays, the positive sign in Eq.~\eqref{eq:Istatic0} corresponds to the portion of the trajectory that approaches the black hole, while the negative sign corresponds to the portion moving away from the black hole.

Figs.~\ref{fig:shadow5} and~\ref{fig:shadow6} present the observed appearance of a de-SBH with static spherical accretions, alongside the observed intensity of the SBH for comparison.  A prominent feature is that the location of the brightest ring, which corresponds to the peak in the observed intensity, is independent of both the size of the black hole (taking into account that the considered de-SBH has a smaller event horizon than the SBH) and the emission models used. The brightest ring is precisely located at the apparent radius of the photon sphere. In all the emission models, the observed intensity outside the brightest ring for the de-SBH is nearly identical to that for the SBH. However, a difference might emerge in the observed intensity within the region inside the brightest ring. Specifically, the de-SBH may appear slightly brighter than the SBH. This is due to the fact that light rays have a longer arclength for the de-SBH compared to the SBH, as the de-SBH has a smaller radius for its event horizon. For a slow decreasing emission profile [cf. the left panel of Fig.~\ref{fig:shadow5}], this difference in the brightness could be negligible due to the redshift factor. However, the difference becomes more apparent as the radial emission decreases more rapidly, see the top panel of Fig.~\ref{fig:shadow5} and the left panel of Fig.~\ref{fig:shadow6}.  

\subsection{Free infalling spherical emissions}
\begin{figure}[htbp] 
  \centering 
  \begin{subfigure}[t]{0.3\textwidth}
    \includegraphics[width=\textwidth]{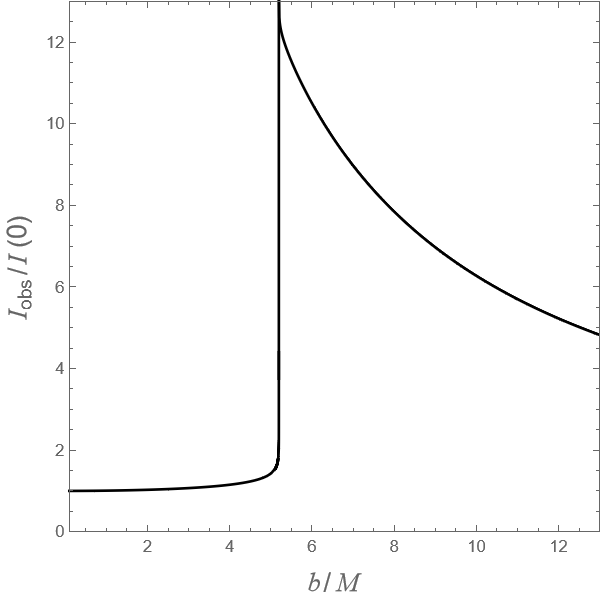}
    \caption*{$j_{1}=1/r^2$}
    \label{fig:datashpereacc5}
  \end{subfigure} \hspace*{2.0mm}
  \begin{subfigure}[t]{0.3\textwidth}
    \includegraphics[width=\textwidth]{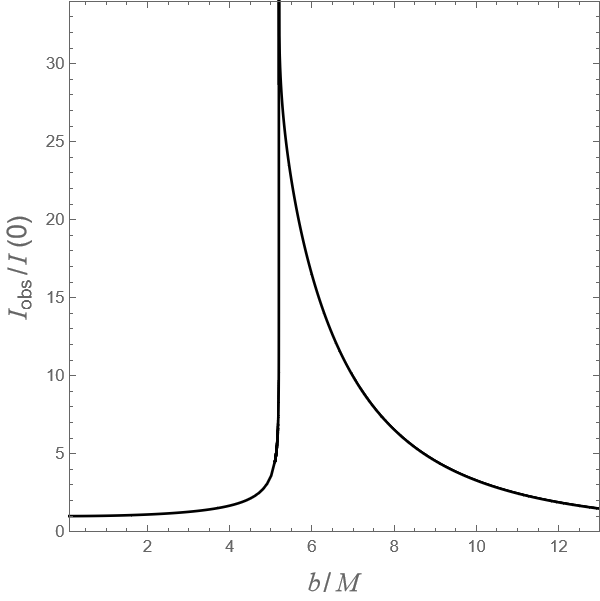}
    \caption*{$j_{1}=1/r^4$}
    \label{fig:datashpereacc6}
  \end{subfigure} \hspace*{2.0mm}
  \begin{subfigure}[t]{0.3\textwidth}
    \includegraphics[width=\textwidth]{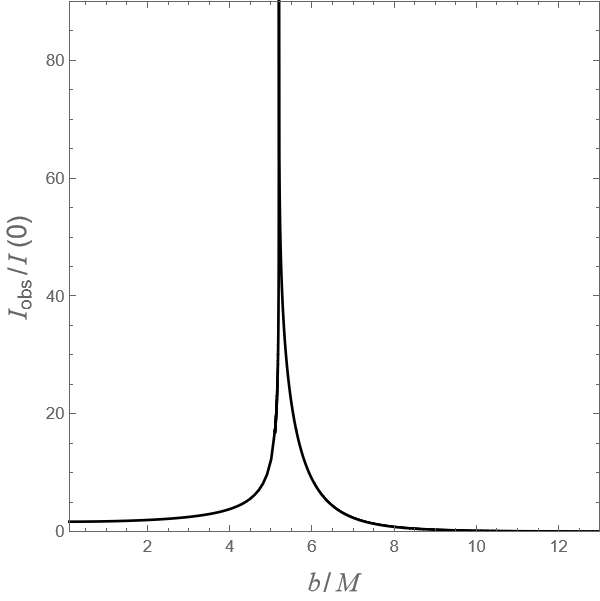}
    \caption*{$j_{1}=1/r^8$}
    \label{fig:datashpereacc7}
  \end{subfigure} \\ \hspace*{6.0mm}
  \begin{subfigure}[t]{0.3\textwidth} 
    \includegraphics[width=\textwidth]{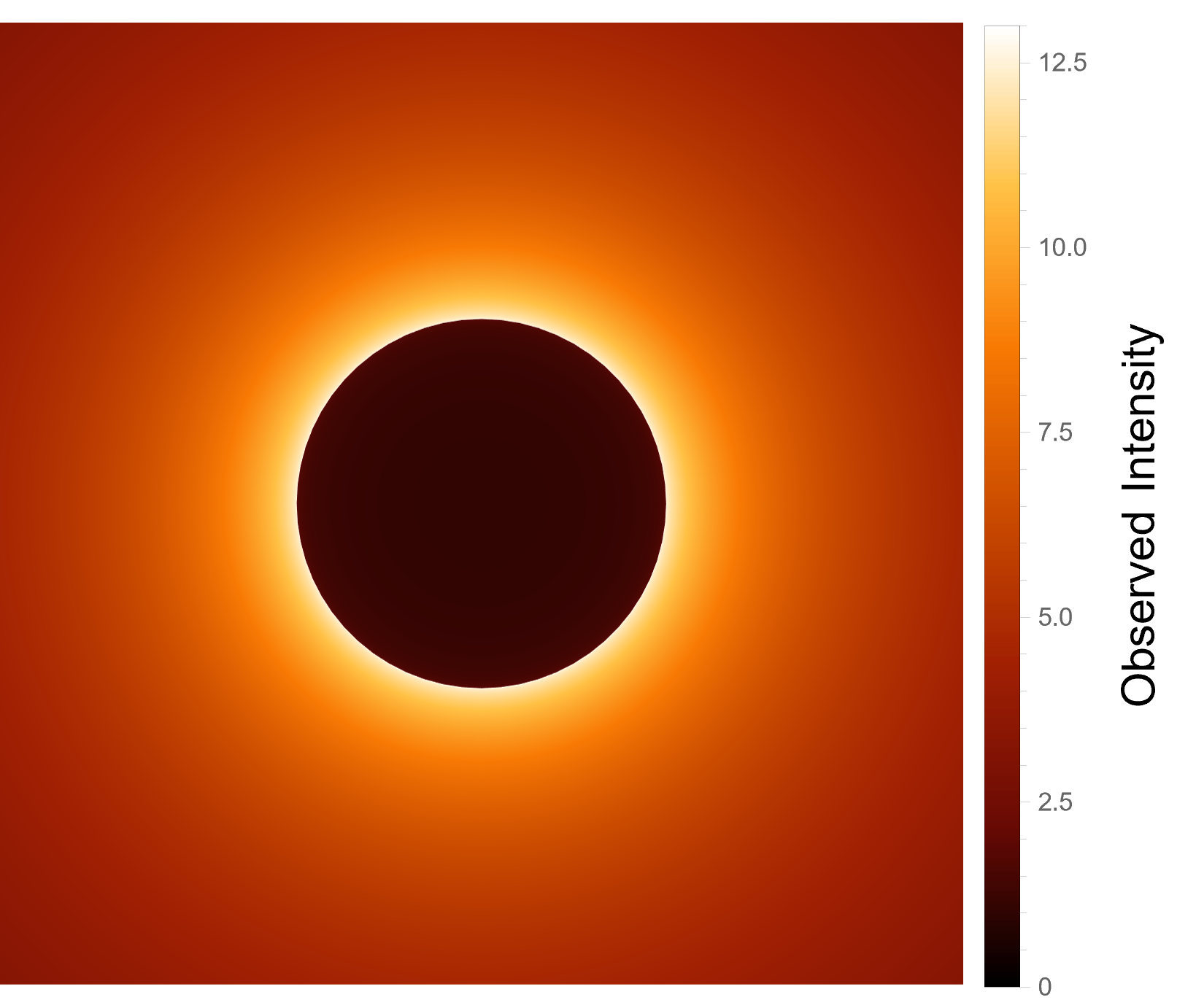}
    \label{fig:dssBHshadowsphere5}
  \end{subfigure} \hspace*{2.0mm}
  \begin{subfigure}[t]{0.3\textwidth} 
    \raisebox{0.0\height}{\includegraphics[width=\textwidth]{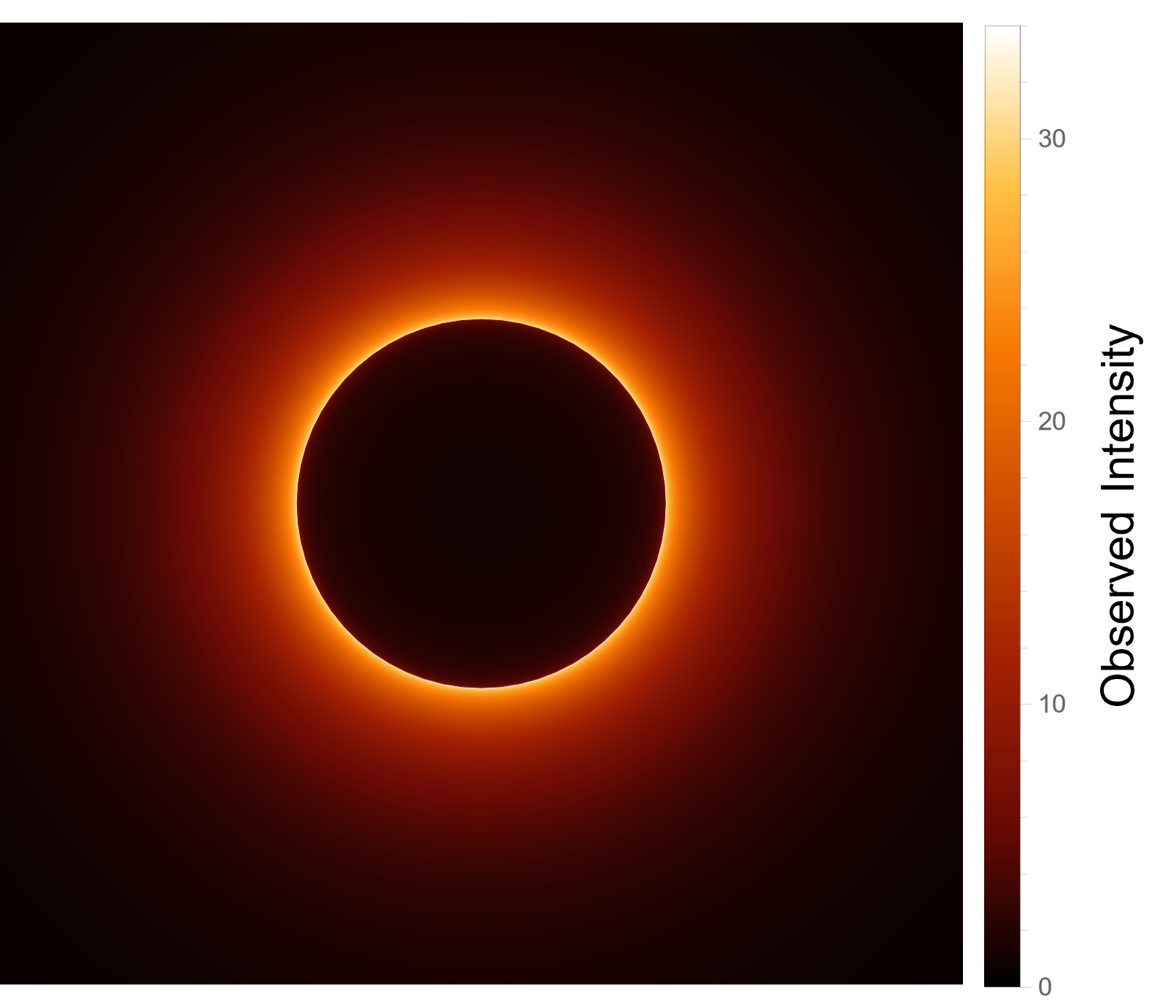}}
    \label{fig:dssBHshadowsphere6}
  \end{subfigure}\hspace*{2.0mm}
  \begin{subfigure}[t]{0.3\textwidth} 
    \raisebox{0.0\height}{\includegraphics[width=\textwidth]{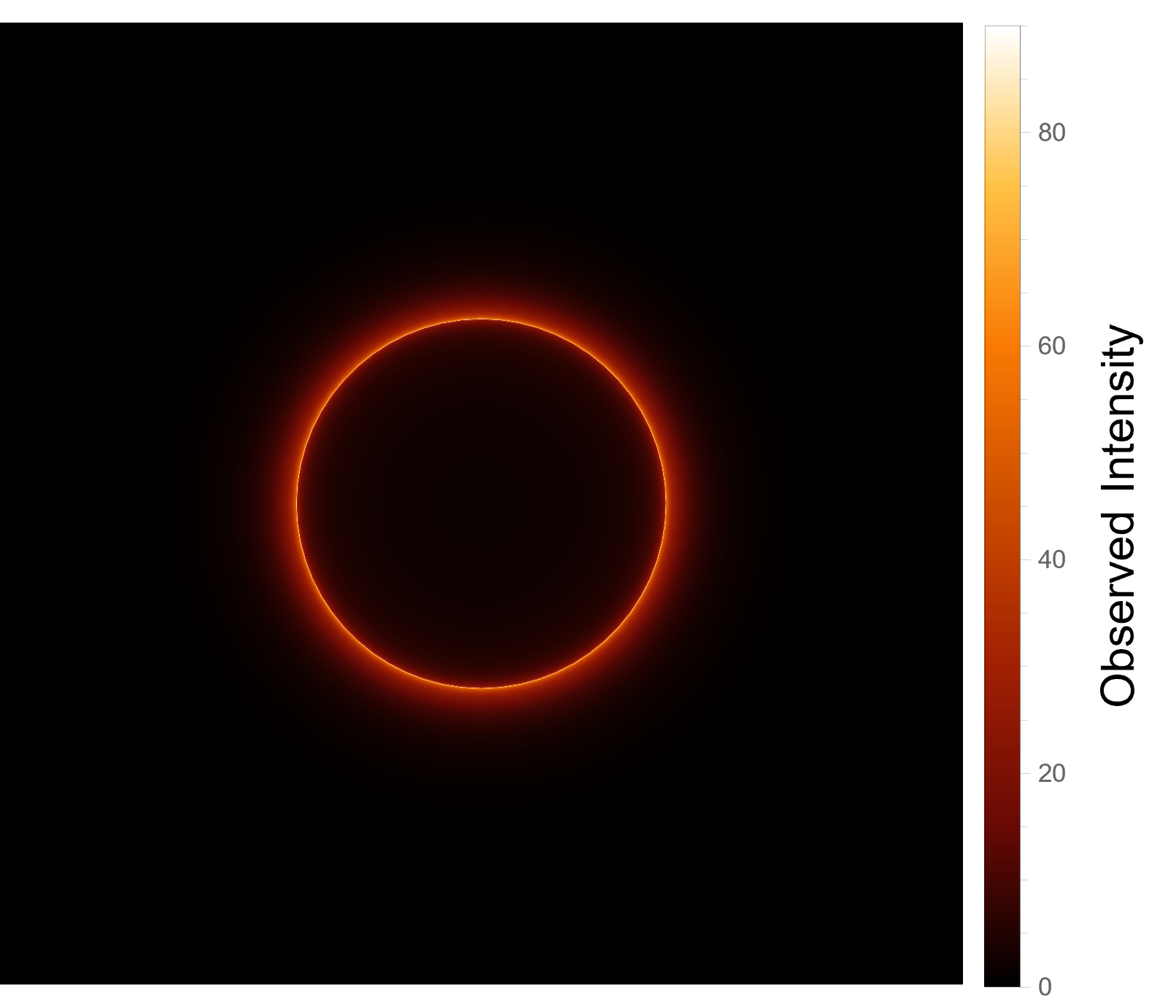}}
    \label{fig:dssBHshadowsphere7}
  \end{subfigure}
  \caption{Observed appearance of a de Sitter-Schwarzschild black hole ($r_g/r_0=1.76$) with radial free infalling spherical accretions. Top panel: Observed intensity $I_{\rm obs}/I(0)$ as a function of the impact parameter $b$, where $I(0)$ indicates the observed intensity at $b=0$. Bottom panel: Optical image of the de Sitter-Schwarzschild black hole. The radial profile of emission from left to right is $j_{1}=1/r^2$, $j_{2}=1/r^4$ and $j_{4}=1/r^8$, respectively. }
  \label{fig:shadow7}
\end{figure}

For free infalling spherical emissions, we have the four-velocity of the emitter
\begin{align}
  u^{\alpha}_{\rm em}=\left(-1/{g_{tt}},-\sqrt{1+g_{tt}},0,0\right)\,.
\end{align}
And now, the redshift factor is given by
\begin{align}
  g=\frac{-g_{tt}}{1\pm \sqrt{1+g_{tt}}\sqrt{1+g_{tt}b^2/r^2}}\,.
\end{align}
It is worth noting that for a light ray emitted at the position closed to the event horizon,  $g_{tt} \to 0$ and the redshift factor can be approximate as $g\approxeq -g_{tt}/2$. Therefore, for the emission closed to the black hole, the redshift factor of the free infalling emission approaches zero more rapidly compared to that of static emission. 

The total observed intensity can be written as
\begin{align}\label{eq:Istatic}
  I_{\rm obs}^{\rm infalling}=\pm \int _{\rm ray} \frac{r\, g^3 \,j_l}{\sqrt{r^2+b^2g_{tt}}}\dd r\,.
\end{align}
The numerical result for this integral is presented in the top panel of Figure~\ref{fig:shadow7}, accompanied by the corresponding optical image of the de-SBH depicted in the bottom panel of the same figure. Similar to the case of static emissions, the location of the brightest ring remains independent of the emission models and precisely positioned at the apparent radius of the photon sphere. However, the relative observed intensity inside the brightest ring for infalling emissions is lower compared to that for static emissions. This discrepancy arises due to the distinct behavior of the redshift factor, as discussed previously. Additionally, in terms of the observed appearance, the de-SBH is nearly indistinguishable from the SBH for the infalling emission models. This is primarily because the redshift factor at small radii is extremely small, making the difference in the size of the black hole negligible.

\section{Conclusion and discussion}
\label{sec:conclusion}

Nonsingular black holes are of utmost importance as they provide an alternative approach to addressing the singularity problem inherent in classical black hole solutions.  In this paper, we specifically investigated a well-known model of nonsingular black holes, namely the de Sitter-Schwarzschild black hole (de-SBH)~\cite{Dymnikova:1992ux}, with a primary focus on its optical appearance. 

The radius of the photon sphere for the de-SBH is approximately the same as that of the Schwarzschild black hole (SBH). However, the (external) event horizon of de-SBH is slightly smaller compared to SBH. For instance, in this paper, our primary focus is on a specific de-SBH with a ratio of $r_g/r_0=1.76$, which results in an external event horizon located at approximately $r_h \approx 1.74M$. We found that, in the case of a spherically symmetric accretion model, the outer edge of the shadow observed in the black hole image is independent of the specific details of the accretion flow and even the \emph{size} (event horizon) of the black hole. This result has been explicitly demonstrated in Section~\ref{Optically and geometrically thin spherical emission}: Despite the smaller radius of the event horizon in the de-SBH compared to the SBH, the size of the observed shadow could remain the same. The outer edge of the black hole shadow is precisely located at the apparent radius of the photon sphere, denoted as $b_c$, which is determined solely by the effective potential of null geodesics. This finding emphasizes the robustness and universality of the outer shadow edge in black hole images for spherical emissions, highlighting the fundamental role played by the photon sphere.

For optically and geometrically thin disk emission, the outer edge of the shadow observed in the image from a face-on orientation can vary for different emission models and may also depend on the size of the black hole. In scenarios where there is no emission occurring inside the photon sphere, the radius of the shadow is influenced by the specific details of the emission profile rather than the radius of the event horizon. In this situation, it is not possible to differentiate between the de-SBH and SBH based on their optical appearance. On the other hand, when the emission extends all the way to the event horizon, the radius of the shadow is approximately located at the apparent position of the horizon, which is approximately $b_h^{\rm de-SBH}\approx 2.6M$ for the de-SBH and $b_h^{\rm SBH}\approx 2.9M$ for the SBH. However, it is important to note that despite the small difference in the size of the dark region between the  de-SBH and SBH, this distinction is virtually indistinguishable due to the low brightness at the edge of the shadow in comparison to the overall appearance.  Without a measurement that provides a sufficiently high angular resolution, it becomes challenging to differentiate between these two types of black holes based solely on the size of the shadow.  

Under certain circumstances, it is indeed possible to differentiate between the de-SBH and the SBH based on their optical appearance.  The distinction primarily arises from the behavior of brightness, which varies depending on the emission model. When considering disk emission with a sharp peak near the event horizon, one can examine the overall brightness of the image. As demonstrated in Figures~\ref{fig:shadow3} and~\ref{fig:shadow4}, the de-SBH appears slightly darker than the SBH. This distinction becomes more pronounced as the radial emission decreases more rapidly. Conversely, for static spherical emission characterized by the radial profile $j_l=1/r^{2l}$, one can observe the brightness of the ``shadow" within the brightest ring. The ``shadow" of the de-SBH appears slightly brighter than that of the SBH in this scenario, as illustrated in Figures~\ref{fig:shadow5} and~\ref{fig:shadow6}. This distinction also becomes more noticeable as the radial emission decreases more rapidly.

Real accretion flows in astrophysical systems are complex and can be classified into different categories, including cold and hot accretion flows~\cite{Yuan:2014gma}. These flows involve a wide range of physical processes that are more intricate than the simplified emission models considered in this paper. However, in many galactic nuclei, accretion disks are known to be geometrically thick and quasi-spherical~\cite{Narayan:2019imo}. These disks exhibit properties that lie between the idealized spherical model and the geometrically thin disks studied in this paper. 
As a result, our study emphasizes that differentiating between the de-SBH and the SBH should prioritize evaluating the brightness of the optical image rather than focusing exclusively on the shadow's size. Particularly, noticeable differences in brightness between the de-SBH and the SBH become more apparent when the accretion flow extends closer to the event horizon and exhibits a more rapid decline in emission. Our study would provide a possible examination of the de-SBH and enhance our understanding of nonsingular black holes. 

In conclusion, we would like to highlight two additional points. Firstly, our study focused specifically on the optical appearance of the de-SBH with a ratio of $r_g/r_0=1.76$. In cases where the ratio is larger, the (external) horizon of the de-SBH becomes closer to that of the SBH. Consequently, the differences in the optical appearance between these two types of black holes become less pronounced. Secondly, it is crucial to emphasize that our conclusions regarding the optical appearance of the de-SBH and its distinguishability from the SBH may have broader implications. These findings offer a potential direction for research on black hole models that share the same radius of the photon sphere but vary in horizon radius compared to the SBH. In these instances, differences in optical appearance might not be predominantly determined by variations in the shadow's size but, instead, by disparities in the overall brightness of the image. 

\section*{ACKNOWLEDGEMENTS}

It is a pleasure to thank Xiao-Yan Chew and Yong-Zhuang Li for useful discussions. This work is supported by the Natural Science Research Project of Colleges and Universities in Jiangsu Province (21KJB140001) and Natural Science Foundation of Jiangsu Province (BK20220642). 

\begin{appendix}

\section{Review of the de Sitter-Schwarzschild spacetime}
\label{sec:Reveiw of de Sitter-Schwarzschild spacetime}
\begin{figure}
    \centering
    \includegraphics[scale=1]{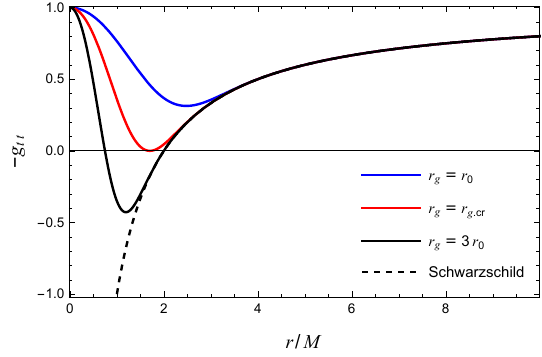}
    \caption{(color online). ${t-t}$ component of de Sitter-Schwarzschild metric for different ratio $r_g/r_0$. }
    \label{fig:figmetric}
  \end{figure}

The de Sitter-Schwarzschild spacetime was considered to be a solution of Einstein field equations with the following density profile~\cite{Dymnikova:1992ux} 
\begin{align}
    \rho(r)=\frac{3}{8\pi r_0^2} \exp\left(-\frac{r^3}{r_0^2 r_g}\right) \,.
\end{align}
One could define the mass function as 
\begin{align}
    \mathcal{M}(r)=4\pi \int_{0}^{r} \rho (x)x^2\dd x\,,
\end{align}
with which the metric \eqref{eq:metric_dssBH} could be written as
\begin{align}
    -g_{00}=g_{11}^{-1}= 1-2\mathcal{M}/r\,.
\end{align}
Depending on the ratio of $r_g/r_0$, the metric \eqref{eq:metric_dssBH} can describe either a black hole or a gravitational object without an event horizon. In the case where there is no event horizon, the object is referred to as a G-lump, a term coined by Dymnikova~\cite{Dymnikova:2003vt}. For a given $r_0$, the critical value of $r_{g.\rm{cr}}$ could be found by solving the following equations
\begin{subequations}\label{eq:app_noname1}
    \begin{align}
        g_{00}(r_{\rm cr})=0 \,,\\
        g_{00}'\Big| _{r=r_{\rm cr}}=0\,,
    \end{align}
\end{subequations}
where the prime denotes the derivative with respect to $r$ and where $r_{\rm cr}$ is location of the horizon. Eq.~\eqref{eq:app_noname1} reduce to 
\begin{subequations}\label{eq:app_noname2}
    \begin{align}
        r_{\rm cr}^2/r_0^2&=1/3+\ln (3r_{\rm cr}^2/r_0^2) \,,\\
        r_{g.\rm{cr}}&=\frac{r_{\rm cr}^3/r_0^3}{\ln (3r_{\rm cr}^2/r_0^2)}\,.
    \end{align}
\end{subequations}
The numerical physical solution for this equation yields $r_{\rm cr}\approx 1.49571 r_0$ and $r_{g.\rm{cr}}\approx 1.75759 r_0$. For $r_g>r_{g.\rm{cr}}$, the metric \eqref{eq:metric_dssBH} describes a black hole with two event horizons and these two horizons degenerate  at $r_g=r_{g.\rm{cr}}$. For $r_g<r_{g.\rm{cr}}$, the metric describes the G-lump. Examples of different de Sitter-Schwarzschild metric are shown in Fig.~\ref{fig:figmetric}.

\section{Geodesics of de Sitter-Schwarzschild black hole}
\label{sec:Geodesics of de-SBH}
For a static spherically symmetric metric which has the form:
\begin{align}\label{eq:app_metric}
    \dd s^2=-B(r)\dd t^2 + B(r)^{-1}\dd r^2+r^2(\dd \theta ^2+\sin^2\theta \dd \phi^2)\,,
\end{align}
the static Killing vector field $(\partial/\partial t)^{\mu}$ and the rotational Killing vector field $(\partial/\partial \phi)^{\mu}$ yield the following constants of the motion for geodesics, respectively:
\begin{subequations}\label{eq:app_constant of motion}
    \begin{align}
        E&=-g_{\mu \nu}\left(\frac{\partial}{\partial t}\right)^{\mu}\left(\frac{\partial}{\partial \lambda}\right)^{\nu}\,,\\
        J&=g_{\mu \nu}\left(\frac{\partial}{\partial \phi}\right)^{\mu}\left(\frac{\partial}{\partial \lambda}\right)^{\nu}\,,
    \end{align}
\end{subequations}
where $\lambda$ being the proper time for massive particle or the affine parameter for massless particle. Without loss of generality, we shall restrict our attention to the equatorial geodesics, i.e., $\theta =\pi/2$.

With the help of Eq.~\eqref{eq:app_constant of motion}, the geodesic equations
\begin{align}
   \frac{\dd ^2x^{\mu}}{\dd \lambda ^2}  +\Gamma^{\mu}_{\rho \sigma}\frac{\dd x^{\rho}}{\dd \lambda }\frac{\dd x^{\sigma}}{\dd \lambda }=0
\end{align}
can be simplified to 
\begin{subequations}\label{eq:app_geodesicseq2}
    \begin{align}\label{eq:app_geodesicseq2_1}
        \frac{E}{B(r)}&=\frac{\dd t}{\dd  \lambda} \,,\\ \label{eq:app_geodesicseq2_2}
       \frac{E^2}{2}&=\frac{1}{2}\left( \frac{\dd r}{\dd  \lambda}\right)^2 +\frac{B(r)}{2} \left(\frac{J^2 }{r^2}+N  \right) \,, \\ \label{eq:app_geodesicseq2_3}
       \frac{J}{r^2}&=\frac{\dd \phi}{\dd \lambda}\,,
    \end{align}
\end{subequations}
with constant $N=0$ for a massless particle and $N=1$ for a massive particle.

Note that Eq.~\eqref{eq:app_geodesicseq2_2} has the same form of the equation for a particle with
unit effective mass and energy $E^2/2$ moving in a one-dimensional effective potential
\begin{align}
    V_{\rm eff}=\frac{B(r)}{2} \left(\frac{J^2 }{r^2}+N  \right)\,.
\end{align}
Combining Eqs.~\eqref{eq:app_geodesicseq2_2} and~\eqref{eq:app_geodesicseq2_3}, we obtain
\begin{align}
    \left( \frac{\dd r}{\dd  \phi}\right)^2-\frac{E^2 r^4}{J^2}+B(r)\left(r^2+\frac
    {Nr^4}{J^2}\right)=0\,.
\end{align}

\begin{figure}
  \centering
  \includegraphics[scale=1]{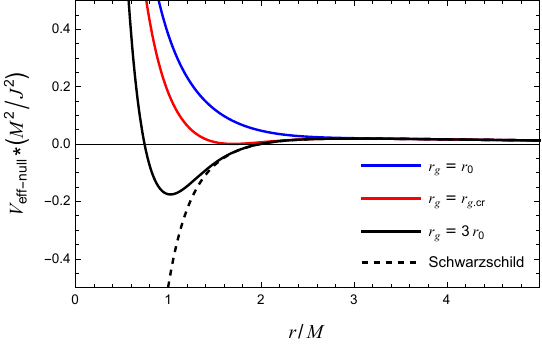}
  \caption{(color online). Effective potential for massless particles in de Sitter-Schwarzschild spacetime with various values of $r_g$. }
  \label{fig:effVnull}
\end{figure}

\begin{figure}
  \centering
  \begin{subfigure}[b]{0.45\textwidth}
    \includegraphics[width=\textwidth]{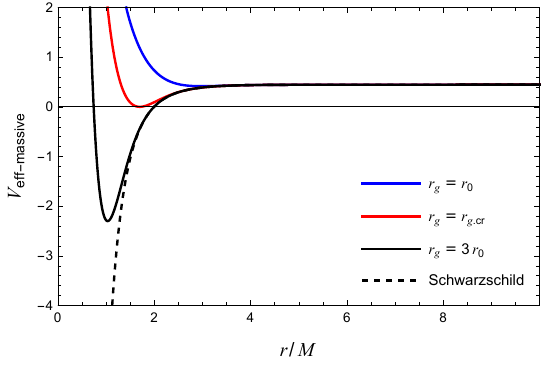}
    \label{fig:figVmassunitM}
  \end{subfigure} 
  \begin{subfigure}[b]{0.45\textwidth}
    \includegraphics[width=\textwidth]{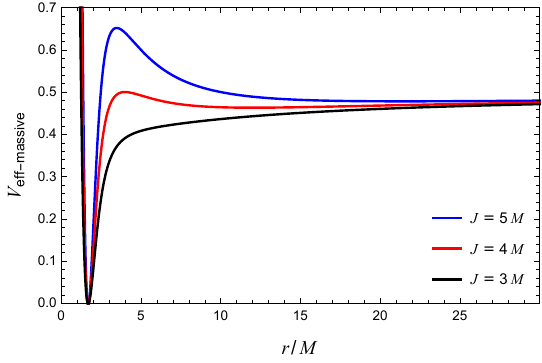}
    \label{fig:figVmassL}
  \end{subfigure}
  \caption{(color online). Left panel: Effective potential for massive particles with angular momentum $J=\sqrt{12}M$ in de Sitter-Schwarzschild spacetime, considering various values of $r_g/r_0$. For comparison purposes, we also include the corresponding results for the Schwarzschild spacetime, represented by dashed line. All these effective potentials have a local minimal at $r\approxeq 6M$, which corresponds to a stable circular orbits. Right panel: Effective potential for massive particles with different values of angular momentum $J$ in a de-SBH with $r_g/r_0=1.76$.}
  \label{fig:figVmass}
\end{figure}

For null geodesics ($N=0$), the effective potential is given by 
\begin{align}
    V_{\rm eff-null}=\frac{J^2 B(r) }{2r^2}\,.
\end{align} 

For a de-SBH, we have 
\begin{subequations}
    \begin{align}
        B(r)=1-\frac{r_g}{r} \left(1-{\rm{e}} ^{-r^3/r_0^2 r_g}\right)\,.
    \end{align}
The equation for the effective potential and ${\dd r}/{\dd  \phi}$ are given by 
    \begin{align}
        V_{\rm eff-null} =& \frac{J^2}{2r^2}-\frac{J^2r_g}{2r^3}+\frac{J^2r_g}{2r^3}{\rm{e}} ^{-r^3/r_0^2 r_g}\,,
        \\
        0=&\left( \frac{\dd r}{\dd  \phi}\right)^2-\frac{E^2 r^4}{J^2}+r^2-r_g r\left(1-{\rm{e}} ^{-r^3/r_0^2 r_g}\right)\,, 
    \end{align}   
\end{subequations}
which are Eqs.~\eqref{eq:effeV} and~\eqref{eq:photon_orbit}, respectively.  Fig.~\ref{fig:effVnull} show the effective potential for massless particles in de Sitter-Schwarzschild spacetime.

For timelike geodesics, $N=1$, and the effective potential of a de-SBH is given by 
\begin{align}
  V_{\rm eff-massive} =& \frac{J^2}{2r^2}-\frac{J^2r_g}{2r^3}+\frac{J^2r_g}{2r^3}{\rm{e}} ^{-r^3/r_0^2 r_g}+\frac{1}{2}-\frac{r_g}{2r}+\frac{r_g}{2r}{\rm{e}} ^{-r^3/r_0^2 r_g}\,.
\end{align}
Examples of the effective potential are plotted in Fig.~\ref{fig:figVmass}. The location of the stable circular orbits can be denoted by $r_{\rm SCO}$ ($r_{\rm SCO}>r_{h}$), which corresponds to the local minimal of the effective potential outside the event horizon. Note that $r_{\rm SCO}$ is $J$ dependent, and the minimal $r_{\rm SCO}$ gives the location of the innermost stable circular orbits $r_{\rm ISCO}$. The innermost stable circular orbits for de-SBHs is nearly equal to that of the SBH, i.e., $r_{\rm ISCO}\approxeq 6M$.

\end{appendix}
\bibliography{reference}

\end{document}